\documentclass[epj]{svjour}
\pdfoutput=1
\usepackage{amsmath}

\newcommand{\beq}{\begin{equation}}
\newcommand{\eeq}{\end{equation}}

\newcommand{\carbon}{\rm ^{12}C}
\newcommand{\deuteron}{\rm ^{2}H}

\newcommand{\D}{\displaystyle}

\usepackage{array, graphicx, subfigure}
\usepackage{graphics}
\begin{document}
\title{Effective Spectral Function for Quasielastic Scattering on Nuclei}

\author{A. Bodek\inst{1},   M. E. Christy\inst{2}, B. Coopersmith\inst{1},}
\institute{Department of Physics and Astronomy, University of
Rochester, Rochester, NY  14627-0171 USA
\and Hampton University, Hampton, Virginia, 23668 USA}

\date{Received: date /  Revised }
\date{Received: date /- version 11.0 September 15,  2014 }
%
\abstract{
Spectral functions that are used in  neutrino event generators 
 to model quasielastic (QE)  scattering from nuclear targets include  Fermi gas,  Local Thomas Fermi gas (LTF),  Bodek-Ritchie Fermi gas with high momentum tail, and the Benhar-Fantoni two dimensional spectral function.  We find that the $\nu$ dependence of  predictions of these spectral functions for the  QE  differential cross sections (${d^2\sigma}/{dQ^2 d\nu}$)
   are in disagreement with the prediction of the  $\psi'$ superscaling function which  is extracted from fits to  quasielastic electron scattering data on nuclear targets.  It is known that spectral functions do not fully describe quasielastic scattering because they only model the initial state.  Final state interactions distort the shape of  
 the differential cross section at the peak and  increase the cross section  at the tails of the distribution.  We show that  the  kinematic distributions predicted by  the $\psi'$ superscaling formalism  can be well described  with a modified    {\it {effective spectral function}} (ESF).   By construction, models using ESF in combination with  the transverse enhancement contribution correctly predict  electron QE  scattering data.
 }  
{ 
}
\maketitle
\section{Introduction}
\label{intro}
Neutrino oscillation experiments make use of neutrino Monte Carlo (MC) event generators to model the  cross sections and  kinematic distributions of the leptonic and hadronic final state of neutrino interactions on nuclear targets.   Therefore, reliable simulations of the effects of  Fermi motion and other nuclear effects are important.  In order to model neutrino cross sections we need to model the vector part, the  axial-vector part, and axial-vector interference.  Because of the conservation of the vector current (CVC), the same models should be able to  reliably predict the  QE electron scattering cross section on nuclear targets. Unfortunately, none of  the models which are currently implemented in neutrino MC generators are able to do it.  In this paper, we propose an approach which guarantees agreement with  quasielaststic (QE)  electron  scattering data by construction.

\subsection{QE Scattering from independent nucleons}

\begin{figure}
\begin{center}
\includegraphics[width=1.6in,height=1.5in]{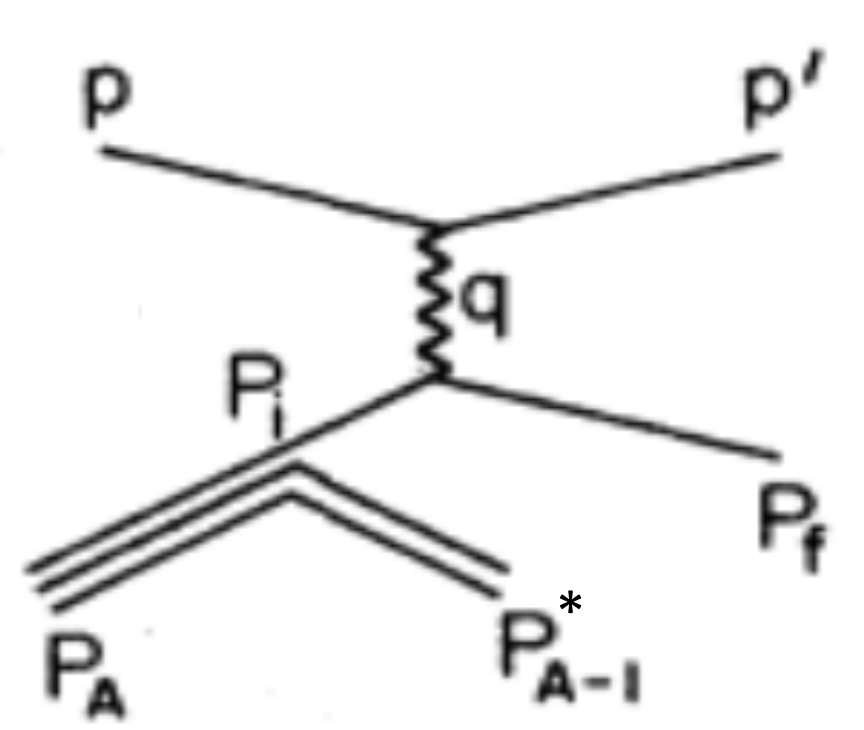}
\includegraphics[width=2.in,height=1.7in]{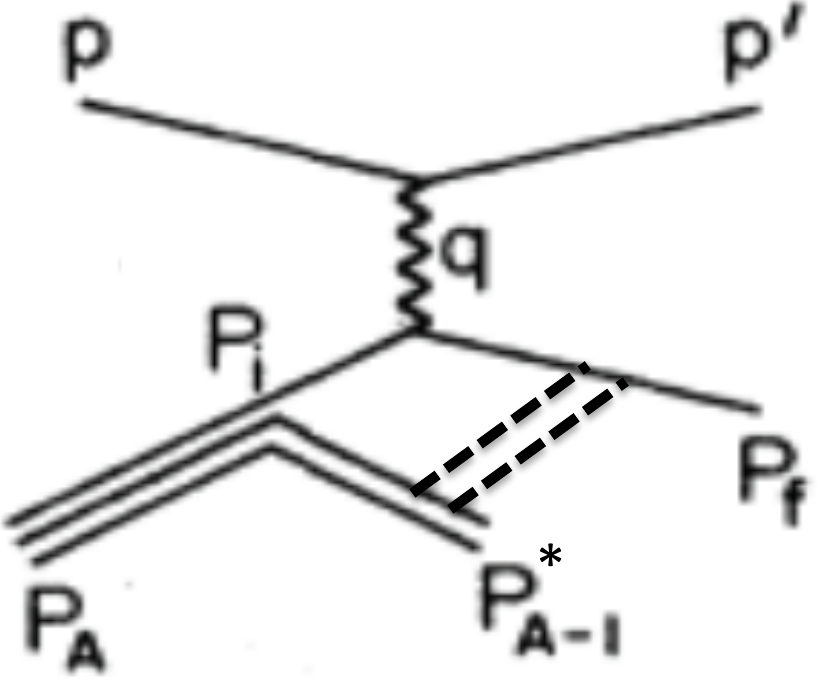}
\caption{ Top: Scattering from an off-shell bound neutron of momentum $\bf{P_i}=k$ in 
a nucleus of mass A.  The on-shell recoil  $[A-1]^*$ (spectator) nucleus has a momentum $\bf{P_{A-1}^*= P_s=-k}$. This process
is referred to as the 1p1h process (one proton one hole).
 Bottom: The 1p1h process including final state interaction (of the first kind) with another nucleon.  
}
\label{Aoff-shell}
\end{center}
\end{figure} 
%

\begin{figure}
\begin{center}
\includegraphics[width=3.5in,height=2.5in]{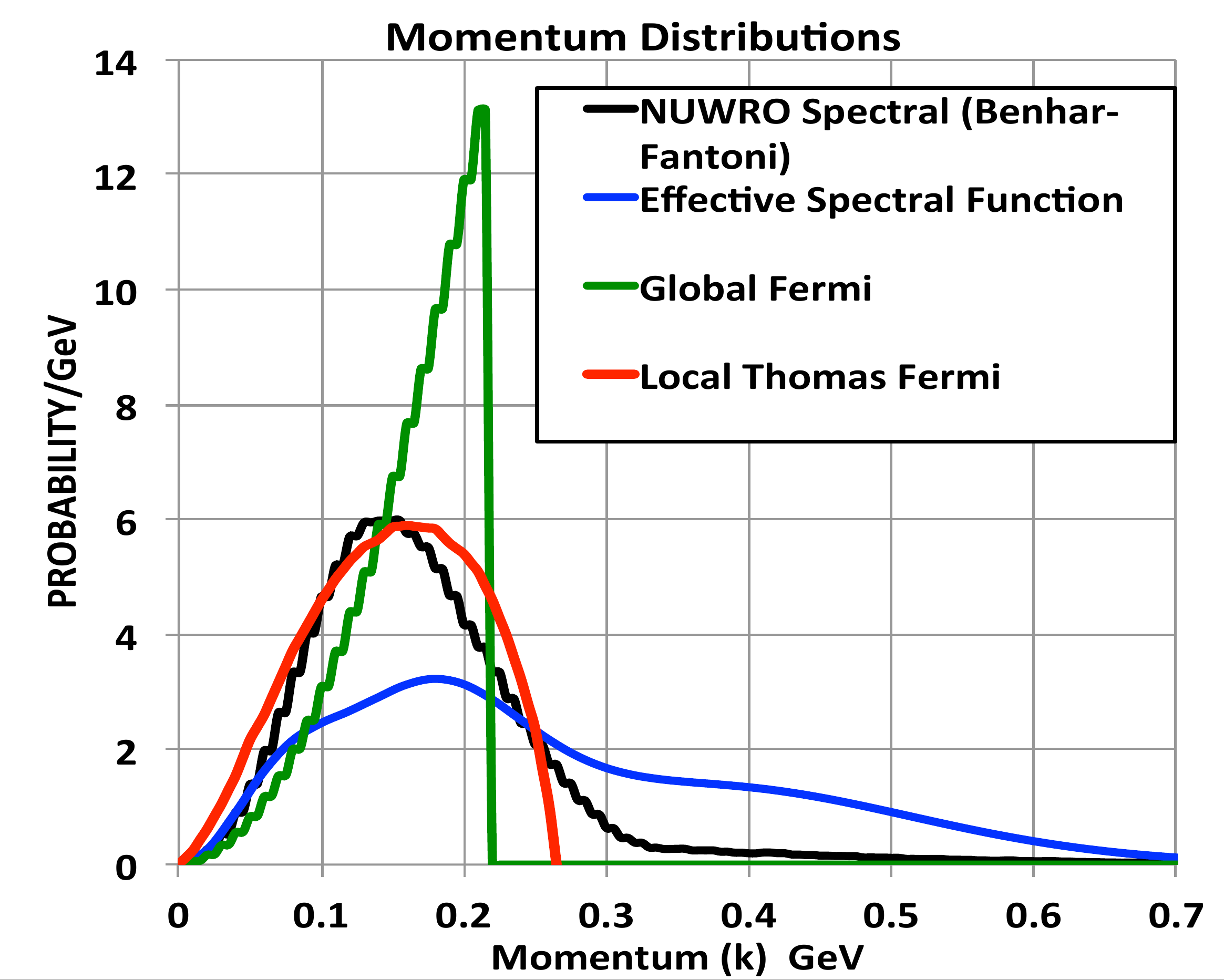}
\caption{ 
Nucleon momentum distributions  in a $\carbon$ nucleus for several spectral functions. The curve labeled "Global Fermi" gas is
the momentum distribution for the Fermi gas model (equation \ref{eqFermi} in Appendix B).
The blue line is the momentum distribution for the  {\it {effective spectral function}} described in this paper (color online).
}
\label{momentum}
\end{center}
\end{figure}

The  top panel of Fig. \ref{Aoff-shell} is the general diagram for QE lepton (election, muon or neutrino)  scattering from a nucleon which is bound in a nucleus of mass $M_A$.  In this paper, we focus on charged current neutrino scattering.
The scattering is from an off-shell bound neutron of momentum $\bf{P_i}=k$.  The on-shell recoil  $[A-1]^*$ (spectator) nucleus has a momentum $\bf{P_{A-1}^*= P_s=-k}$. This process
is referred to as the 1p1h process (one proton one hole).
The * is used to indicate that the spectator nucleus is not in the ground state because it has one hole.
The four-momentum transfer to the nuclear target is defined as $q = (\vec k,  \nu)$.  Here  $\nu$ is the energy transfer, and
$Q^2= -q^2 = \nu^2- \vec{q}^2$ is the square of the four-momentum transfer.  For free nucleons the energy transfer $\nu$ is
equal to $Q^2/2M_N$ where $M_N$ is the mass of the nucleon. At a fixed value of $Q^2$,  QE scattering  on nucleons bound in a nucleus yields a distribution in $\nu$ which peaks at  $\nu=Q^2/2M_N$.  In this communication, the term
"normalized quasielastic distribution" refers to the normalized  differential cross section $\frac{1}{\sigma} \frac{d\sigma}{d\nu}(Q^2,\nu)=\frac{d^2\sigma /dQ^2 d\nu} {<d\sigma/dQ^2>} $ where $<\frac{d\sigma}{d Q^2}>$ is  the integral of $[\frac{d^2\sigma}{dQ^2 d\nu } ]d\nu$ over all values of $\nu$ (for a given value of $Q^2$).

The  bottom  panel of Fig. \ref{Aoff-shell} shows the same QE lepton scattering process, but now also  including  a final state interaction  with another nucleon in the scattering process. This final state interaction modifies
the scattering amplitude and  therefore can change the kinematics of the final state lepton. In this paper, we refer to it as 
"final state interaction of the first kind" (FSI).

The final state nucleon can then undergo more  interactions with other nucleons
in the spectator nucleus.  These  interactions do not change the energy of the final state lepton.  We refer to this
kind of final state interaction as "final state interaction of the second kind".
Final state interactions  of the second kind reduce the energy of the final state nucleon.

\subsection{Spectral functions}

 In general, neutrino event generators assume that the  scattering occurs on  independent nucleons which are bound in the nucleus.
Generators such as  GENIE\cite{genie}, NEUGEN\cite{neugen}, NEUT\cite{neut}, NUANCE\cite{nuance} NuWro \cite{nuwro} and GiBUU\cite{gibuu}   account for nucleon binding effects  by modeling the momentum distributions and removal energy of nucleons in nuclear targets. Functions that describe the momentum distributions and removal energy of nucleons from nuclei are referred to as spectral functions.

 Spectral functions can take the  simple form of a momentum distribution and a fixed removal energy (e.g. Fermi gas model\cite{moniz}),  or the more complicated form of a two dimensional (2D)  distribution in both momentum and removal energy (e.g. Benhar-Fantoni spectral function \cite{bf}).  

Fig. \ref{momentum} shows  the nucleon momentum distributions in a $\carbon$ nucleus for some of the
spectral functions that are currently being used. The solid green line is the nucleon  momentum distribution for the  
 Fermi gas\cite{moniz} model (labeled "Global Fermi" gas)  which is currently implemented in all neutrino event generators is given in equation \ref{eqFermi} of Appendix B).
 The solid black line is  the projected momentum distribution of the  Benhar-Fantoni \cite{bf}  2D spectral function as implemented in NuWro.  
The solid red line is the nucleon momentum distribution of the Local-Thomas-Fermi gas (LTF) model\cite{gibuu} which is implemented in NURWO and GiBUU. 

It is known that  theoretical calculations using spectral functions do not fully describe the shape of the  quasielastic  peak for  electron scattering on nuclear targets .  This is  because the calculations  only model the initial state (shown on the top panel of Fig. \ref{Aoff-shell}), and do not account for final state interactions of the first kind  (shown on the bottom panel of Fig. \ref{Aoff-shell}) .
Because FSI changes the amplitude of the scattering,  it modifies the shape of $\frac{1}{\sigma}\frac{d\sigma}{d\nu}$. FSI reduces the cross section at the peak and  increases the cross section  at the tails of the distribution.

In contrast to the spectral function formalism, predictions using the  $\psi'$ superscaling formalism\cite{super1,super2} fully describe the longitudinal response function of  quasielastic electron scattering data on nuclear targets.  This is expected since the calculations use a   $\psi'$ superscaling function  which is  directly extracted from the  longitudinal  component of measured electron scattering quasielastic differential cross sections.
 
However,  although  $\psi'$ superscaling provides a very good description of the final state lepton in
QE scattering,  $\psi'$ superscaling is not implemented as an option in neutrino MC event  generators that are
currently used  neutrino experiments.  There are  specific technical issues  that are associated
with implementing any  theoretical model within the framework of a MC generator. In
addition,  $\psi'$ superscaling
does  not provide a detailed description of the composition of the hadronic final state.  Therefore, it must
also be combined with other models to include details about the composition of the  hadronic final state. 

Because the machinery to 
model both the leptonic and hadronic final state for various spectral functions is  already implemented 
in all neutrino MC generators,  adding another spectral function as an option can be implemented in a few days.
In this communication  we present the parameters for a new {\it {effective spectral function}}  that reproduces the  
 kinematics of the final state lepton  predicted by $\psi'$ superscaling.  The momentum distribution for this ESF for  $\carbon$ is shown as the  blue line in Fig. \ref{momentum}.
\begin{figure}
\begin{center}
\includegraphics[width=3.5in,height=3.5in]{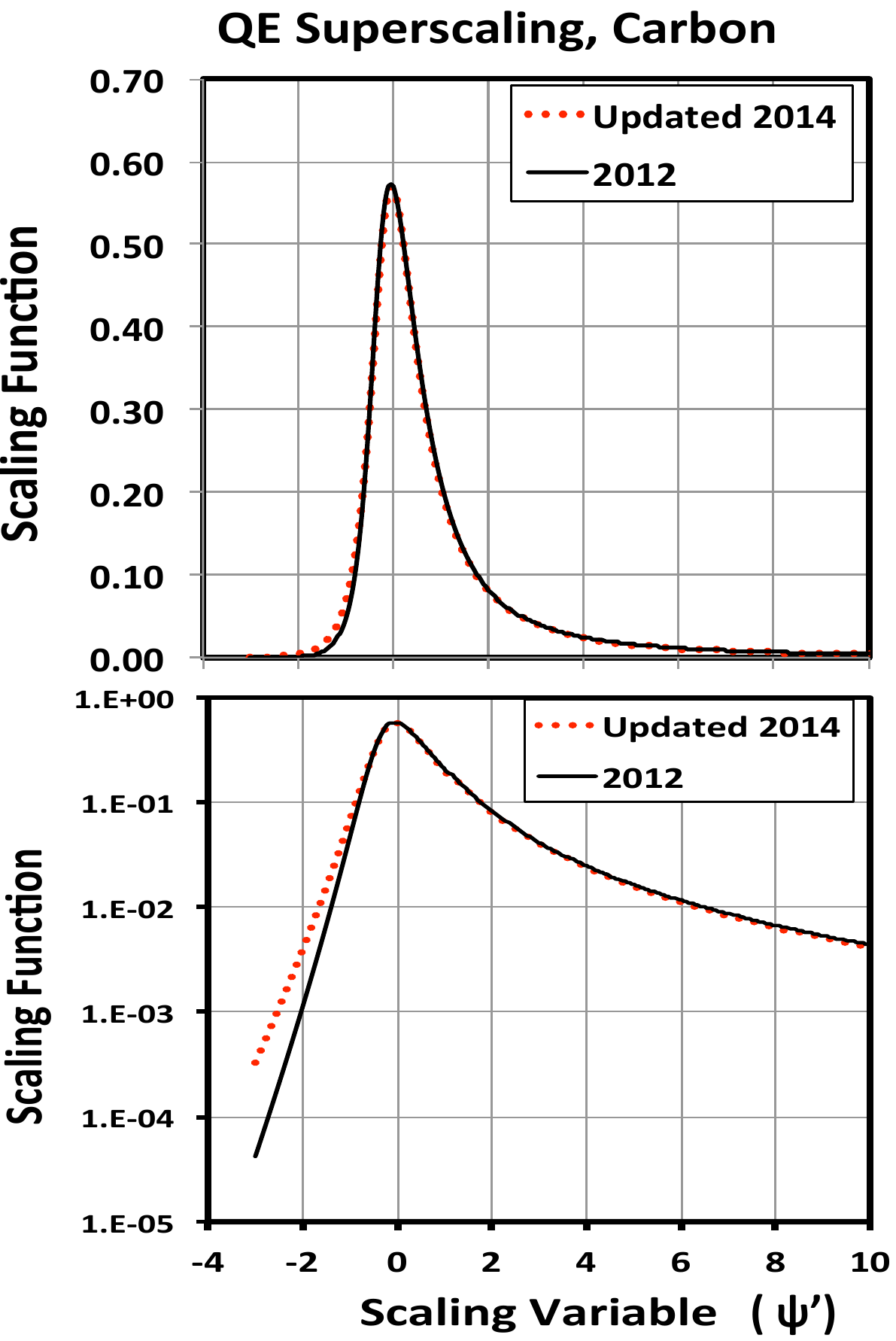}
\caption{  The  $\psi'$ superscaling distribution extracted from a fit to electron scattering data used by  Bosted and Mamyan \cite{super2} (solid black line labeled as 2012), and the  superscaling function  extracted from a more recent updated fit \cite{eric}  to data from a large number of quasielastic  electron scattering experiments  on $\carbon$ (dotted red line labeled as 2014). The panel on top shows the superscaling functions on a  a linear scale and the panel on the bottom shows the same superscaling functions on  a logarithmic scale. The integral of the curve has been normalized to unity (color online).
}
\label{superfunction}
\end{center}
\end{figure} 

\subsection { The $\psi'$ superscaling functions for QE scattering}
 
The  $\psi$ scaling variable\cite{super1,super2} is defined as:
\begin{equation}
\psi\equiv \frac{1}{\sqrt{\xi_F}} \frac{\lambda-\tau}{\sqrt{(1+\lambda)\tau+
\kappa\sqrt{\tau(1+\tau)}}},
\label{eq:psi}
\end{equation} 
where  $\xi_F\equiv \sqrt{1+\eta_F^2}-1$,  $\eta_F \equiv K_F/M_n$, $\lambda \equiv\nu /2M_n$, $\kappa \equiv {|\vec q| }/2M_n$ and $\tau \equiv|Q^{2}|/4M_n^{2}=\kappa ^{2}-\lambda ^{2}$. 

The  $\psi'$ superscaling variable includes
a correction that accounts for the removal energy from the nucleus. This is achieved by
replacing  $\nu$ with $\nu-E_{\mathrm{shift}}$, which  forces the maximum of the QE response to occur at 
$\psi^\prime=0$.
This is equivalent to taking $\lambda\to\lambda'=
\lambda-\lambda_{\mathrm{shift}}$ with $\lambda_{\mathrm{shift}}=E_{\mathrm{shift}}/2M_n$ and
correspondingly $\tau\to\tau'=\kappa^2-\lambda'^2$ in eq.~(\ref{eq:psi}). 
QE scattering on all nuclei (except for the deuteron) is described using the same universal superscaling function. The only
parameters which are specific to each nucleus are   the Fermi broadening parameter $K_F$ and the  
 energy  shift  parameter  $E_{\mathrm{shift}}$.

  Fig. \ref{superfunction} shows  two parametrizations of  $\psi'$ superscaling functions extracted from quasielastic  electron scattering data  on $\carbon$.  Shown  is the  $\psi'$ superscaling distribution extracted from a fit to electron scattering data used by  Bosted and Mamyan \cite{super2} (solid black line labeled as 2012), and the  superscaling function  extracted from a recent updated fit\cite{eric}  to data from a large number of quasielastic  electron scattering experiments  on $\carbon$ (dotted red line labeled as 2014). The panel on top shows the superscaling functions on a  a linear scale and the panel on the bottom shows the same superscaling functions on  a logarithmic scale.   
   
  The 2014    $\psi'$ superscaling function is given by: 
   \begin{equation}
F(\psi^\prime) = \frac{ 1.3429 }{K_F
[1 + 1.7119^2 (\psi^\prime + 0.19525)^2] 
(1 + e^{-1.69 \psi^\prime})}
\end{equation}
 The 2012   $\psi'$ superscaling function is given by: 
   \begin{equation}
F(\psi^\prime) = \frac {1.5576 }{K_F
[1 + 1.7720^2 (\psi^\prime + 0.3014)^2]
(1 + e^{-2.4291 \psi^\prime})}
\label{F2014}
\end{equation}
For both the 2012 and 2014 parametrizations the values of the Fermi motion parameter $K_F$ and  energy shift  parameter $E_{\mathrm{shift}}$  (given in Table~\ref{tab:kfes}) are taken from  ref. \cite{super2}. 

The  $\psi'$ superscaling function is extracted from the longitudinal QE cross section
for $Q^2>0.3$~GeV$^2$ where there are no Pauli blocking effects. At very 
low values of $Q^2$, the  QE differential
cross sections predicted by the   $\psi'$ superscaling should be multiplied by
a Pauli blocking factor $K_{Pauli}^{nuclei}(Q^2)$
which   reduces the predicted cross sections at low $Q^2$. The Pauli suppression
factor (see Fig. \ref{DCpauli}) is  given\cite{super2} by the function
\begin{equation}
K_{Pauli}^{nuclei} =\frac{3}{4} \frac{|\vec q|}{K_F}(1 - \frac{1}{12} (\frac {|\vec q|}{K_F})^2)
\label{nucpauli}
\end{equation}
For $ |\vec q| < 2K_F$, otherwise no Pauli suppression correction is made.  Here  $ |\vec q| =\sqrt{Q^2+\nu^2}$ is the absolute magnitude of the momentum transfer to the target nucleus,

 In this paper we show that  the  normalized differential quasielastic  cross section    $\frac{1}{\sigma} \frac{d\sigma}{d\nu}(Q^2,\nu)$  predicted by  the $\psi'$ superscaling formalism  can be well described by predictions  of a  modified   {\it {effective spectral function}}  (ESF).    The parameters of
 the ESF are obtained by requiring that the ESF  predictions for    $\frac{1}{\sigma} \frac{d\sigma}{d\nu}(Q^2,\nu)$  at  $Q^2$ values of 0.1,  0.3, 0.5 and 0.7  GeV$^2$ be in agreement with  the predictions  of  the  2014  $\psi'$  superscaling  function  given in eq. \ref{F2014}.
 
  The predictions of the 
 $\psi'$  formalism are given by  
 
  $$\frac{1}{\sigma} \frac{d\sigma}{d\nu}(Q^2,\nu)=   \frac{1}{N} F(\psi^\prime)$$
where N  the integral of $F(\psi^\prime) d\nu $ over all values of $\nu$ (for a given value of $Q^2$).

 %
\begin{table}
\centering
\begin{tabular}{ccccccccc}
\hline
$A$ & $K_F(\psi')$ (GeV) & $E_{\mathrm{shift}}(\psi')$ (GeV) \\ \hline
2 & 0.100 & 0.001 \\
3 & 0.115 & 0.001 \\
$3<A<8$ & 0.190 & 0.017 \\
$7<A<17$ & 0.228 & 0.0165 \\
$16<A<26$ & 0.230 & 0.023 \\
$25<A<39$ & 0.236 & 0.018 \\
$38<A<56$ & 0.241 & 0.028 \\
$55<A<61$ & 0.241 & 0.023 \\
$A>60$ & 0.245 & 0.018 \\
\hline
\hline
\end{tabular}
\caption{Values of Fermi-broadening parameter $K_F$
and  energy shift $E_{\mathrm{shift}}$ used in the   $\psi'$ superscaling prediction for different nuclei.
The parameters for deuterium (A=2) are  to be taken as a crude approximation only, and
deuterium is treated differently as discussed in section \ref{deuteron-sub}.}
\label{tab:kfes}
\end{table}
%
\begin{figure}
\includegraphics[width=3.5in,height=2.5in]{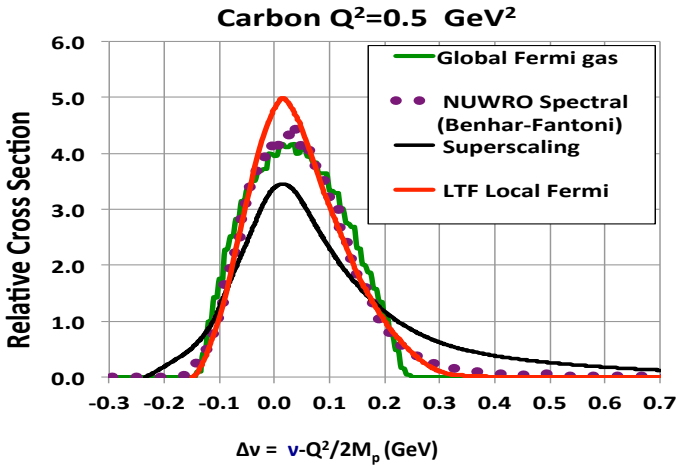}
\caption{ Comparison of the  $\psi'$ superscaling prediction (solid black line)  for  the normalized  $\frac{1}{\sigma} \frac{d\sigma}{d\nu}(Q^2,\nu)$  at $Q^2$=0.5 GeV$^2$ for 10 GeV neutrinos on $\carbon$ 
to the predictions of several spectral function models.  Here   $\frac{1}{\sigma} \frac{d\sigma}{d\nu}(Q^2,\nu)$ 
is plotted versus  $\Delta \nu$
The curve labeled "Global Fermi" gas  is
the distribution for the Fermi gas model given in Appendix B (eq. \ref{nuFermi}).
 The  
  predictions of  the  spectral function models are in disagreement with the predictions
of  $\psi'$ superscaling 
(color online).
}
\label{nuwro-vs-scaling}
\end{figure} 
%
%
\subsection{Comparison of models for quasielastic scattering}
For electron scattering, the nuclear response function is extracted
from  the normalized longitudinal differential 
cross section at a fixed vale of 
$Q^2$  ($\frac{1}{\sigma} \frac{d\sigma}{d\nu}(Q^2,\nu)$).  Here
$\sigma$  is the integral of  $ \frac{d\sigma}{d\nu}$ for  a fixed
value of $Q^2$. The normalization removes  the effects of the $Q^2$-dependent
nucleon vector form factors.
In models which  assume scattering from independent nucleons,
the response functions for the longitudinal and transverse QE cross
sections are the same. 

 For neutrino scattering at high energy,  the QE cross section
is dominated by the structure function $W_2$. Therefore, 
in models which assume scattering from independent nucleons
 the normalized
 cross section  ($\frac{1}{\sigma} \frac{d\sigma}{d\nu}(Q^2,\nu)$)
  at a fixed value of $Q^2$ and high neutrino energy is also equal to the nuclear response
  function.  
 For the neutrino case, the normalization removes the  effects of the $Q^2$-dependent
nucleon vector  and axial form factors.
Fig. \ref{nuwro-vs-scaling}  
shows predictions for the normalized  QE  differential cross sections 
   $\frac{1}{\sigma} \frac{d\sigma}{d\nu}(Q^2,\nu)$ for 10 GeV neutrinos on $\carbon$ at  $Q^2$=0.5 GeV$^2$ for various
   spectral functions.    Here  $\frac{1}{\sigma}\frac{d\sigma}{d\nu}$ 
is plotted versus  $\Delta\nu=\nu-\frac{Q^2}{2M_p}$. 
The prediction of the  $\psi'$ superscaling formalism for   $\frac{1}{\sigma} \frac{d\sigma}{d\nu}(Q^2,\nu)$ is shown  as the solid black line. 
  The solid green line is the prediction using the  "Global Fermi" gas \cite{moniz}
momentum distribution given in Appendix B (eq. \ref{nuFermi}).
 The solid red  line  is the prediction using
the Local  Thomas Fermi gas (LTF)  momentum distribution. 
The dotted purple line is the NuWro prediction using the full two dimensional Benhar-Fantoni\cite{bf}
spectral function.  
 The predictions of all of these spectral functions  for   $\frac{1}{\sigma} \frac{d\sigma}{d\nu}(Q^2,\nu)$  are in disagreement with the predictions
of  the  $\psi'$ superscaling formalism.
%
%
%
\section{Effective Spectral Function for $\carbon$}
%
\subsection{Momentum Distribution}
The probability distribution for a nucleon to have
a momentum $k=|\vec{k}|$ in the nucleus is defined as  
$$P(k) dk=4\pi k^2 |\phi(k)|^2dk.$$
For $k<0.65$ GeV, we  parametrize\cite{bfit} $P(k)$  by the following function:
\begin{eqnarray}
P(k) &=& \frac{\pi}{4c_0}\frac{1}{N}(a_s+a_p+a_t)y^2
\end{eqnarray}
where
\begin{eqnarray}
y &=&\frac{k}{c_0} \nonumber\\
a_s &=&c_1  e^{-(b_sy)^2 } \nonumber\\
a_p &=& c_2(b_py)^2 e^{-(b_py)^2} \nonumber\\
a_t&=& c_3 y^\beta e^{-\alpha(y-2)} \nonumber
\end{eqnarray}
For $k>0.65$ GeV we set $P(k)$ = 0. 
Here,  $c_0=0.197$,  $k$ is in GeV, N is a normalization factor to normalize
the integral of the momentum distribution  from $k$=0 to $k$=0.65 GeV to 1.0,  and P($k$) is in units of  GeV$^{-1}$.
The parameters that describe the projected momentum distribution\cite{bfit}  for  the Benhar-Fantoni\cite{bf} spectral function
for nucleons bound in $\carbon$ 
are  given in the second column of table \ref{fitsC}.  
\begin{table}
\begin{center}
\begin{tabular}{|c|c|c|c|} \hline 
Parameter &  Benhar-Fantoni    &  ESF &  ESF \\ \hline\hline
Nucleus&  $\carbon$   &  $\carbon$  &  $\deuteron$  \\ \hline
$\Delta$ (MeV)  & 2Dspectral        &  12.5   &   0.13   \\ \hline
$f_{1p1h}$  & 2Dspectral        &  0.808   &  0   \\ \hline
$f_{2p2h}$  & 2Dspectral        &  0.192  &  1.00    \\ \hline
$b_s$      &  1.7                &     2.12 &  0.413475    \\ \hline
$b_p$      &  1.77              &  0.7366  &1.75629\\ \hline
$\alpha$  &  1.5                &   12.94  & 8.29029  \\ \hline
$\beta$    &  0.8               &  10.62   & 3.621 x10$^{-3}$   \\ \hline
$c_1$       &  2.823397    &  197.0   & 0.186987\\ \hline
$c_2$       &  7.225905    &  9.94     &  6.24155\\ \hline
$c_3$       & 0.00861524 & 4.36 x10$^{-5}$       & 2.082 x10$^{-4}$\\ \hline
$N$          &  0.985          &  29.64     &  10.33  \\ \hline
\end{tabular}
\caption{ A comparison of the parameters that describe the projected momentum distribution for  the Benhar-Fantoni spectral function for nucleons bound 
in $\carbon$ (2nd column) with  the parameters that describe the {\it{effective spectral function}} (ESF) for $\carbon$ (3rd column).  Here, $\Delta$ is the average binding energy  parameter of the spectator  one-hole nucleus  for the  1p1h process and $f_{1p1h}$ is the fraction of the scattering that occurs via the 1p1h process. 
For the  2p2h process the average binding energy for the two-hole spectator nucleus  is $2\Delta$.  The parameters for the  {\it{effective spectral function}} for deuterium  ($\deuteron$) are given in the 4th column.}
\label{fitsC} 
\end{center}
\end{table}
%

\begin{figure}
\begin{center}
\includegraphics[width=2.in,height=1.7in]{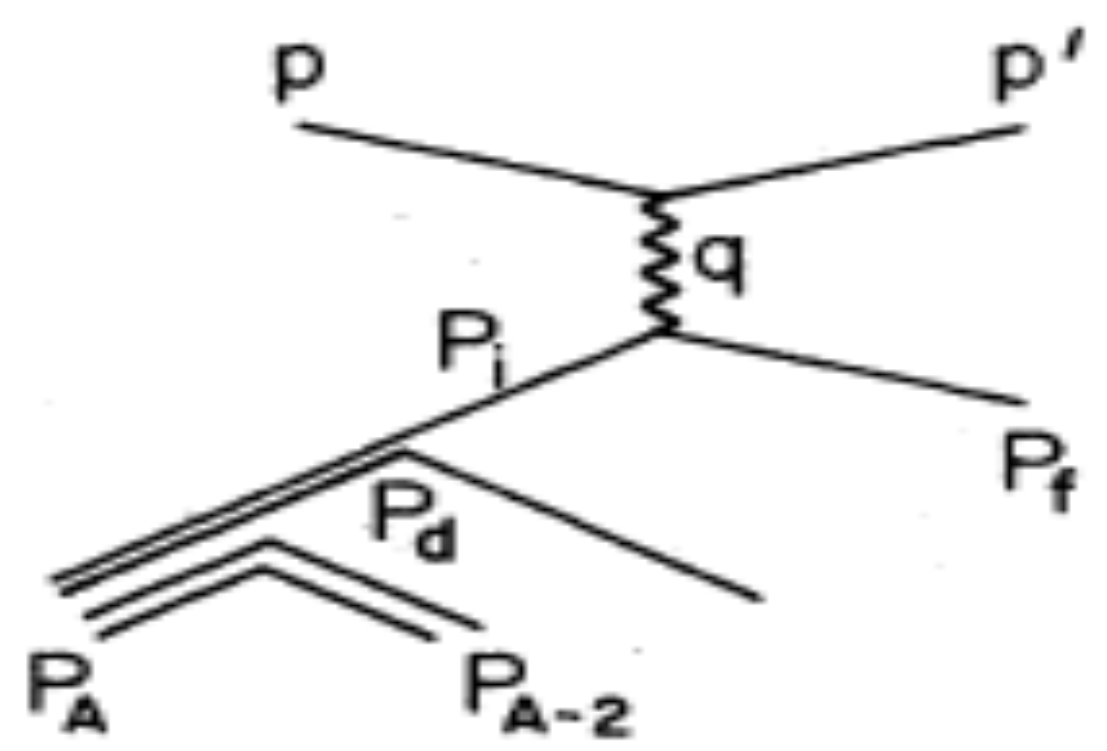}
\caption{ 2p2h process:  Scattering from an off-shell bound neutron of momentum $\bf{P_i=-k}$ from two nucleon
correlations (quasi-deuteron).   The on-shell recoil  spectator nucleon has momentum  $\bf{P_s=k}$.}
\label{Doff-shell}
\end{center}
\end{figure}
%
\begin{figure}
\begin{center}
\includegraphics[width=3.4in,height=2.3in]{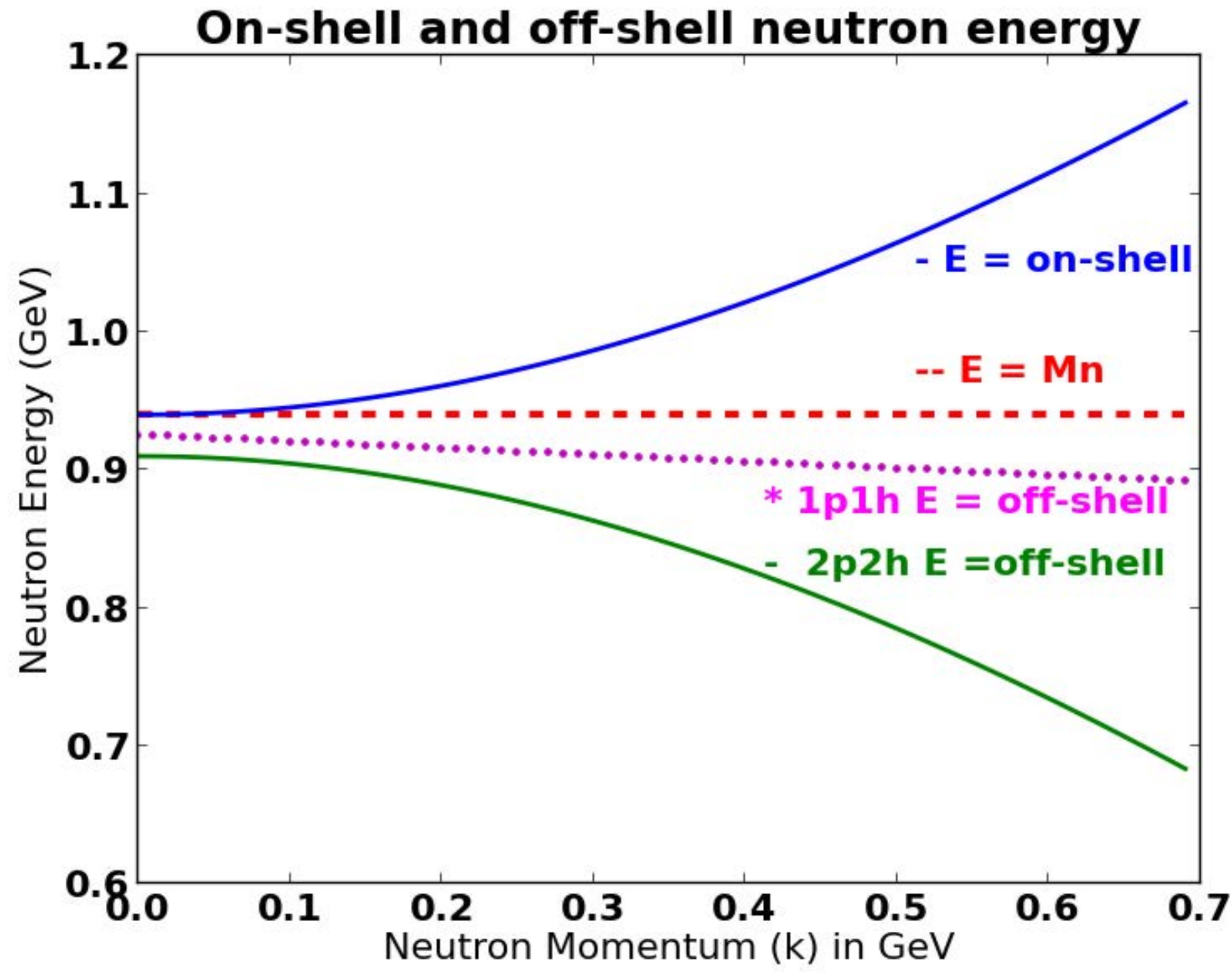}
\caption{ Comparison of energy for on-shell and off-shell bound neutrons in $\carbon$. The on-shell energy is $E_n =\sqrt{k^2+M_n^2}$. The off-shell energy is shown for both the 1p1h ($E_n=M_n-\Delta -\frac{{Vk^2}}{2M_{A-1}^*}$) and 2p2h process ($E_n=(M_p+M_n)-2\Delta-\sqrt{Vk^2+M_p^2}$, where $(M_p+M_n)$  and $\Delta$ is the average binding  energy parameter of the spectator  one-hole nucleus. Shown is the case with V=1. (The factor V is given in equation \ref{Veq} and plotted in  Fig. \ref{vfactor}. V$\approx$1 for $Q^2>$0.3 GeV$^2$) (color online).}
\label{offshell}
\end{center}
\end{figure} 
%
 \begin{figure}
\begin{center}
\includegraphics[width=3.7in,height=2.5in]{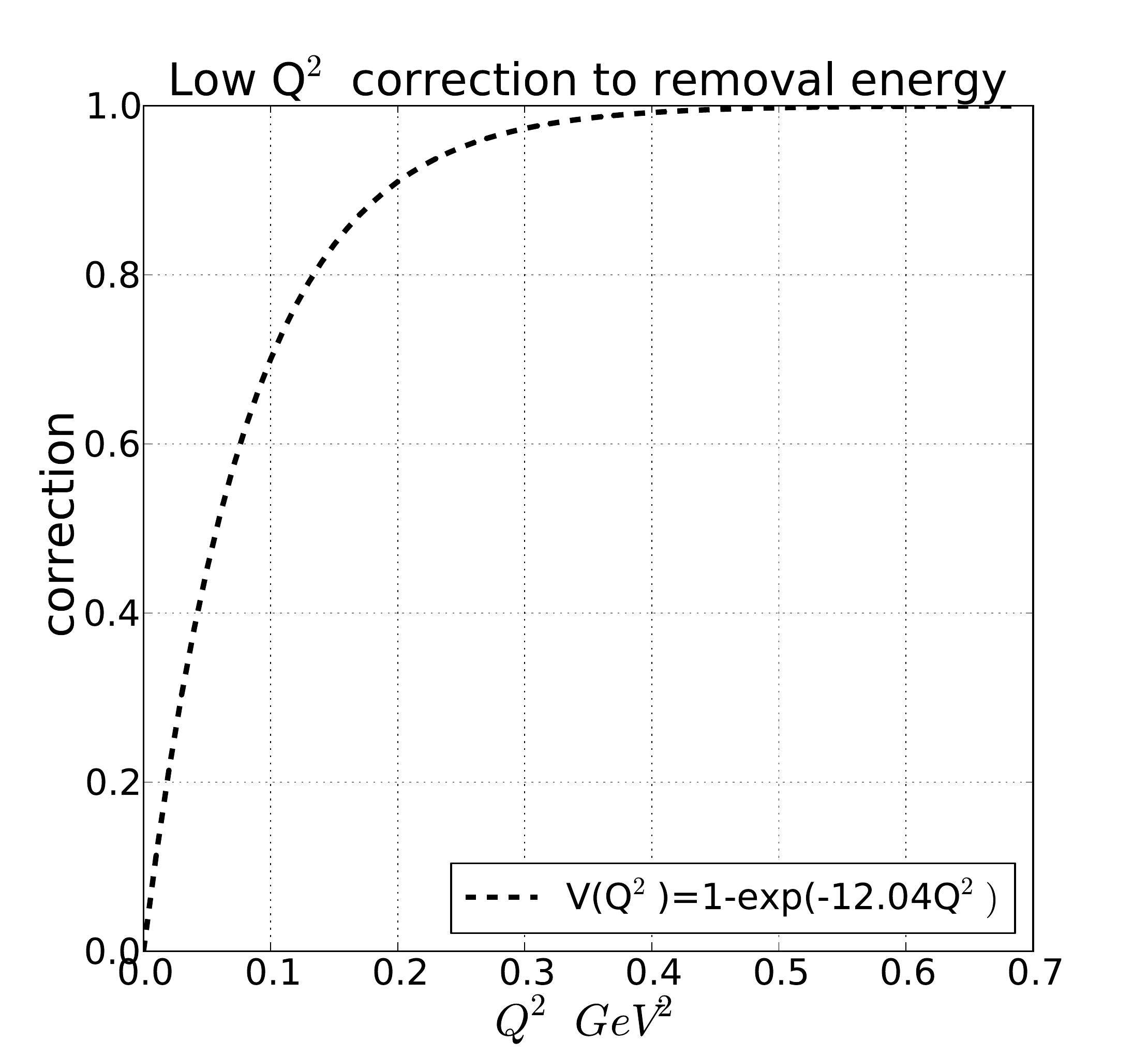}
\caption{The $Q^2$-dependent  correction that accounts for the reduction of the removal
energy at low $Q^2$, e.g. due to final state interaction (of the first kind). }
\label{vfactor}
\end{center}
\end{figure} 
%
%
\subsection{Removal Energy}
The kinematics for neutrino charged current quasielastic 
 scattering from a off-shell bound neutron with momentum
$\bf {k}$ and energy  $E_n$ are given by:
\begin{eqnarray}
\label{eq-nu}
(M_{n}')^2 &= & (E_n)^2 - {Vk^2} \\
M_p^2 &= &(M_{n}')^2 + 2E_n\nu-2|\vec q| k_z- Q^2  \nonumber\\
\nu &= & E_\nu - E_\mu  =   \frac{Q^2+M_p^2- (M_{n}')^2 + 2|\vec q |k_z}{(E_n)} \nonumber\\
V(Q^2) &= &  1 -e^{-xQ^2},~~x=12.04
\label{Veq}
\end{eqnarray}

For scattering from a single off-shell nucleon, the term $V(Q^2)$ multiplying $k^2$  in Equations \ref{eq-nu},  \ref{En1p1h} and  \ref{En2p2h}  (and also Equations   \ref{WWW}, \ref{eq-W1p1h},  and \ref{eq-W2p2h}) should be 1.0.  However, we find that in order to make the spectral function predictions agree with $\psi'$ superscaling at very low $Q^2$ (e.g. $Q^2<0.3~GeV^2$) we need to apply a $Q^2$-dependent correction  to reduce the removal energy, e.g. due to  final state interaction (of the first kind) at low $Q^2$. This factor is given in equation \ref{Veq} and plotted in   Fig. \ref{vfactor}.
 
 The value of the parameter $x$=12.04 GeV$^{-2}$  was extracted from the fits discussed in section \ref{sectionfit}.
As mentioned earlier, $\vec q$ is the  momentum transfer to the neutron. 
We define the component of the initial neutron momentum $\bf k$ which is
parallel to $\vec q$ as $k_{z}$.
The expression for   $E_n$ depends  on the process and is given by  Equations \ref{En1p1h} and  \ref{En2p2h} for the 1p1h, and 2p2h
process, respectively.


We  assume
 that the off-shell energy  ($E_n$)  for a  bound neutron with momentum
 $\bf{k}$ can  only take two possible values\cite{Bodek-Ritchie}. We refer to the 
 first possibility  as the 1p1h process (one proton, one hole  in the final state).
 The second possibility is the 2p2h  process(two protons and two holes in the final state).
 
    In our {\it{effective spectral function}} model the 1p1h process occurs
 with probability $f_{1p1h}$, and the 2p2h process   occurs with
 probability of $1-f_{1p1h}$.  For simplicity, we assume that  the probability  $f_{1p1h}$
 is independent of  the momentum 
 of the bound nucleon.
 \subsubsection{The 1p1h process}
The 1p1h process refers to scattering from an independent neutron in the nucleus resulting in a final state proton and
a hole in the spectator nucleus.
Fig. \ref{Aoff-shell} illustrates the  1p1h process  (for $Q^2>0.3$ GeV$^2$),
for the  scattering from an off-shell bound neutron of momentum $\bf{-k}$ in a nucleus of mass A\cite{Bodek-Ritchie}.   In the 1p1h process, momentum is balanced by an  on-shell recoil  $[A-1]^*$  nucleus  which has momentum $\bf{P_{A-1}^*}=\bf{P_s}=k$ and an average binding energy 
parameter $\Delta$, where  $ M_{A}-M_{A-1}^*=M_n +\Delta$.  The initial state off-shell neutron has energy $E_n$ which is given by:
\begin{eqnarray}
E_n (1p1h)&= & M_A - \sqrt{Vk^2+(M_{A-1}^*)^2} \nonumber \\
&\approx& M_n-\Delta-\frac{Vk^2}{2M_{A-1}^*}
\label{En1p1h}
\end{eqnarray}
The final state includes a proton and an $[A-1]^*$ nucleus which is in an excited state because the removal of the nucleon
leaves  hole in the energy levels of the nucleus. 

 For the 1p1h process, the removal energy of a nucleon includes the following two contributions: 
\begin{itemize}
\item The binding energy parameter  $\Delta$  where $ M_{A}-M_{A-1}^*=M_n +\Delta$.
\item  The  kinetic energy of the recoil spectator nucleus  $\frac{Vk^2}{2M_{A-1}^*}$.
\end{itemize}

 \begin{figure}
\begin{center}
\includegraphics[width=3.5in,height=2.5in]{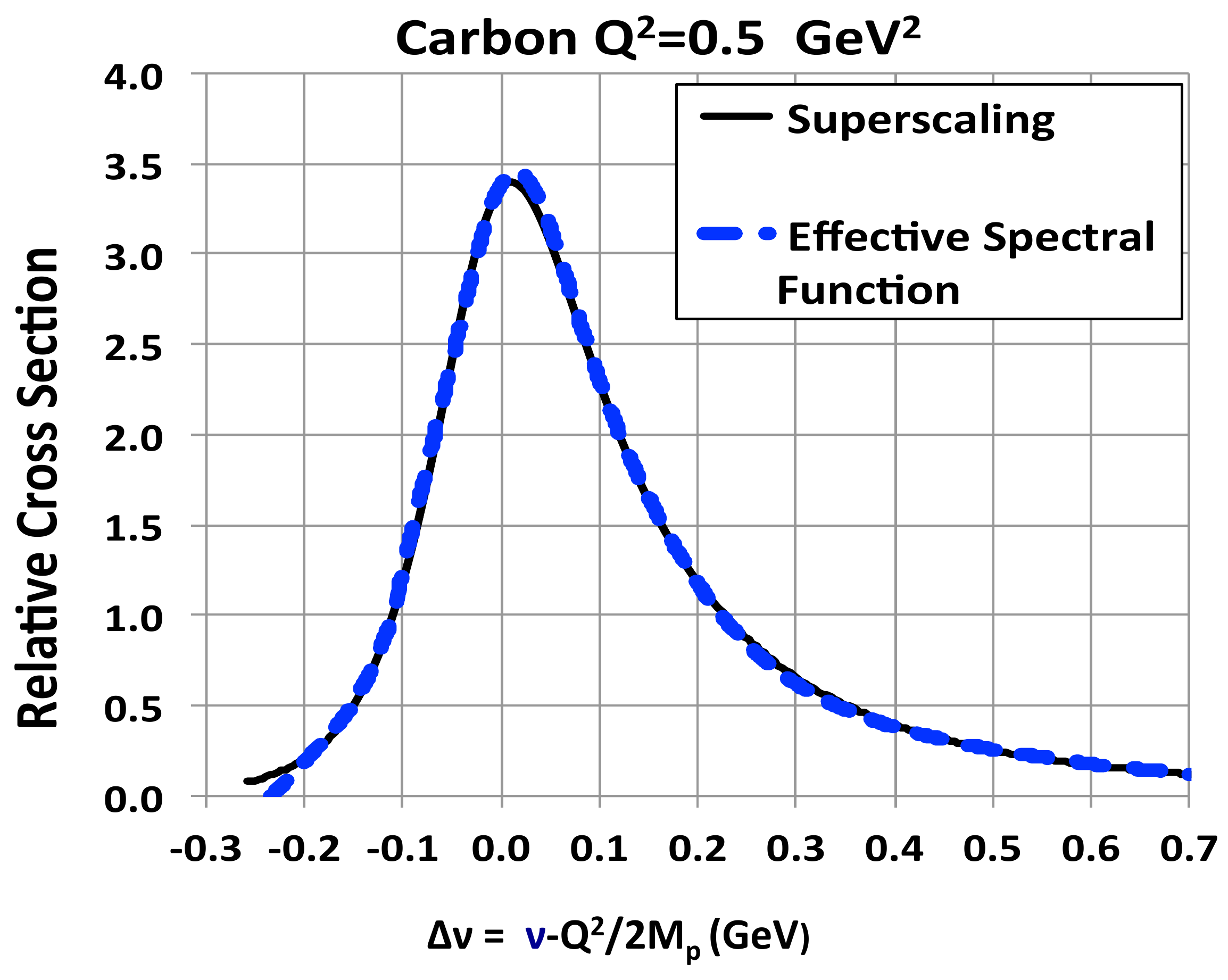}
\caption{ Comparison of the prediction for the normalized QE differential cross section   ($\frac{1}{\sigma} \frac{d\sigma}{d\nu}(Q^2,\nu)$)   for $\carbon$
from the  {\it{effective spectral function}}
to the prediction of   $\psi'$ superscaling. The predictions 
are shown as
a function of $\Delta \nu$ at  $Q^2$=0.5 GeV$^2$.  The prediction of the
effective spectral function are calculated from equation \ref{integral} in Appendix B.
For $Q^2$=0.5 GeV$^2$  the prediction
of the  {\it{effective spectral function}} are almost identical to the prediction of  $\psi'$  superscaling (color online). 
}
\label{finaleffective}
\end{center}
\end{figure} 
%


  \subsubsection{ The 2p2h process}
In general, there are are several processes which result in two  (or more) nucleons and a spectator  excited nucleus with two (or more)  holes in final state:
\begin{itemize}
\item  Two nucleon correlations in initial state (quasi deuteron) which are  often  referred to as short range correlations (SRC).
\item  Final state interaction  (of the first kind) resulting in a larger energy transfer to the hadronic final state (as modeled by superscaling).
\item  Enhancement of the transverse cross sections ("Transverse Enhancement") from meson exchange currents
(MEC) and isobar excitation.
\end{itemize}

In the {\it{effective spectral function}}  approach the lepton energy spectrum  for all three processes is modeled as originating from the  two nucleon correlation process. This accounts for the additional energy shift resulting from the removal of two nucleons from the nucleus.

 Fig. \ref{Doff-shell} illustrates the  2p2h process for  scattering from an off-shell bound neutron of momentum $\bf{-k}$ (for $Q^2>0.3$ GeV$^2$).
The momentum of the interacting nucleon in the initial state
 is balanced by a single on-shell correlated recoil  nucleon  which has momentum $\bf{k}$. The $[A-2]^*$ spectator nucleus is left
 with two holes. The initial state off-shell neutron has energy $E_n$ which is given by:
\begin{eqnarray}
E_n (2p2h)&= & (M_p+M_n) - 2\Delta -\sqrt{Vk^2+M_{p}^2} 
\label{En2p2h}
\end{eqnarray}
\noindent where  $V$  is given by eq. \ref{Veq}.

For the 2p2h process, the removal energy of a nucleon includes the following two contributions: 
\begin{itemize}
\item The binding energy parameter  $2\Delta$  where $ M_{A}-M_{A-2}^*=M_n+M_p +2\Delta$.
\item  The  kinetic energy of the recoil spectator nucleon given by  $\sqrt{Vk^2+M_{p}^2} $.
\end{itemize}

Fig. \ref{offshell} shows a comparison of the total energy for on-shell and off-shell bound neutrons in $\carbon$ as a function of neutron momentum k (for $Q^2>0.3$ GeV$^2$ where V$\approx$1.0). The energy for an unbound  on-shell neutron is $E_n =\sqrt{Vk^2+M_n^2}$. The off-shell energy of a bound neutron  is shown for
both the 1p1h ($E_n=M_n-\Delta-\frac{\bf{Vk^2}}{2M_{A-1}^*}$) and the  2p2h process ($E_n=(M_p+M_n)-2\Delta-\sqrt{Vk^2+M_p^2}$).  

In the  {\it{effective spectral function}}  approach,  all effects of final state interaction (of the first kind)  are absorbed
in the initial state {\it {effective spectral function}}.
 The parameters of
 the {\it {effective spectral function}} are obtained 
 by finding the parameters  $x$, 
$\Delta$, $f_{1p1h}$,  $b_s$, $b_p$, $\alpha$,  $\beta$, $c_1$, $c_2$, $c_3$   and $N$   
  for which the predictions of the     {\it {effective spectral function}}  best describe 
the predictions of  the $\psi'$ superscaling formalism for $(1/\sigma)d\sigma/d\nu$  at  $Q^2$ values of 0.1, 0.3, 0.5 and 0.7  GeV$^2$. 

Fig. \ref{finaleffective} compares predictions for  $\frac{1}{\sigma} \frac{d\sigma}{d\nu}(Q^2,\nu)$  for $\carbon$
as  a function of $\Delta\nu$ at $Q^2$=0.5 GeV$^2$.
The prediction of  the   {\it{effective spectral function}} is the dashed blue curve.
The prediction of the    $\psi'$ superscaling model is the solid black curve. 
For $Q^2$=0.5 GeV$^2$  the prediction
of  the  {\it{effective spectral function}} is almost identical to the prediction of  $\psi'$  superscaling.
All of the  prediction for the
effective spectral function are calculated from equation \ref{integral} in Appendix B.

For the 2p2h process, each of the two final state nucleons can also undergo final state interactions  (of the second kind) with other nucleons
in the spectator $[A-2]^*$ nucleus. 
%
\begin{figure}
\begin{center}
\includegraphics[width=3.5in,height=3.5in]{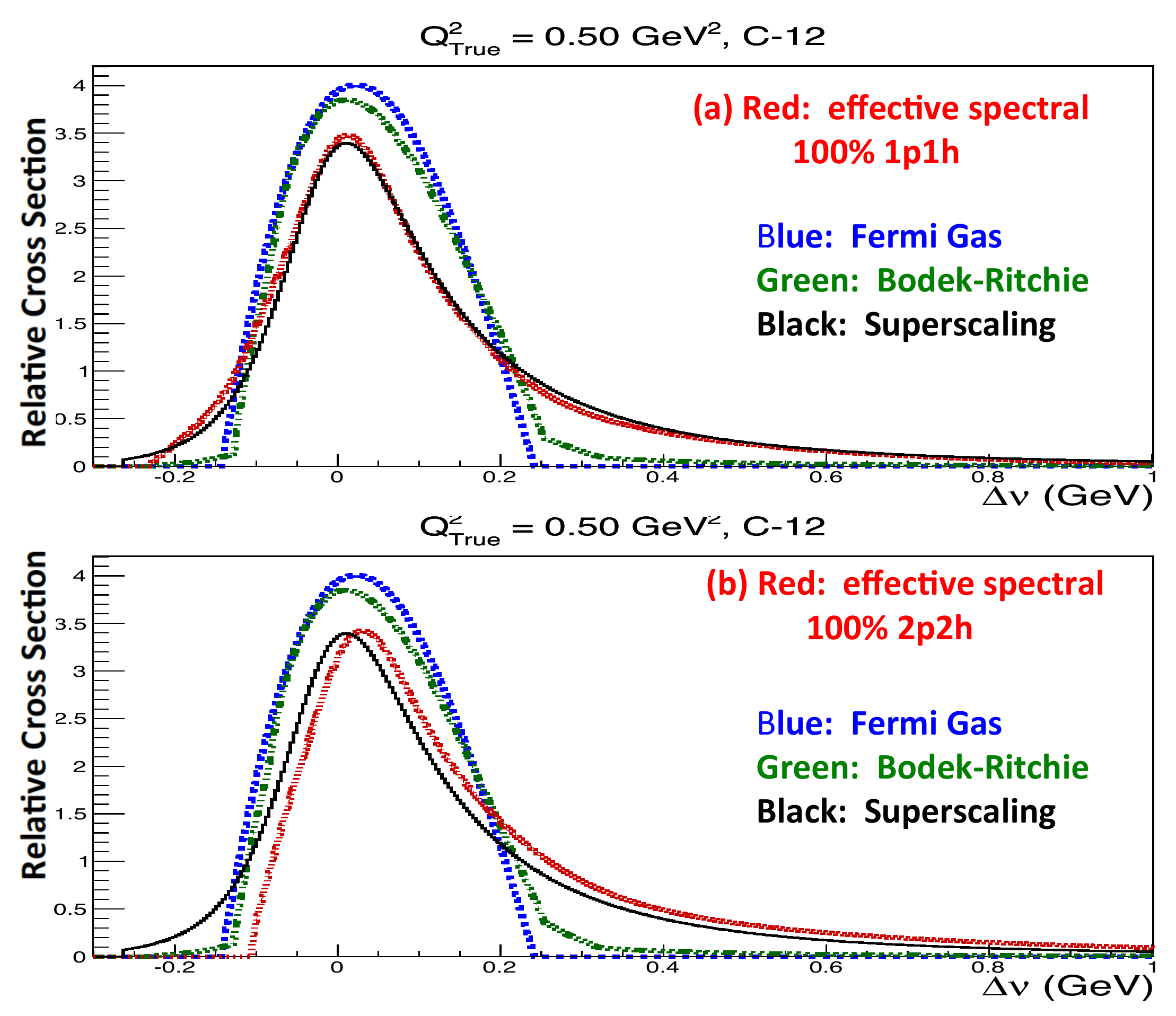}
\caption{Comparison of the  two contributions to the normalized QE cross section ($\frac{1}{\sigma} \frac{d\sigma}{d\nu}(Q^2,\nu)$)  for $\carbon$
from the {\it{effective spectral function}}. (a) For the 1p1h component, (b) For the 2p2h
component.  Here, each contribution (shown in red) is  normalized to 1.0. Also shown (for reference) 
are the predictions for superscaling (black line), for the Fermi Gas model (blue), and the predictions from the  Bodek-Ritchie\cite{Bodek-Ritchie} Fermi gas model which includes a high momentum
contribution from two nucleon correlations (green) (color online). }
\label{1p1h2p2h}
\end{center}
\end{figure} 
%
%
\begin{figure}
\begin{center}
\includegraphics[width=3.5in,height=1.35in]{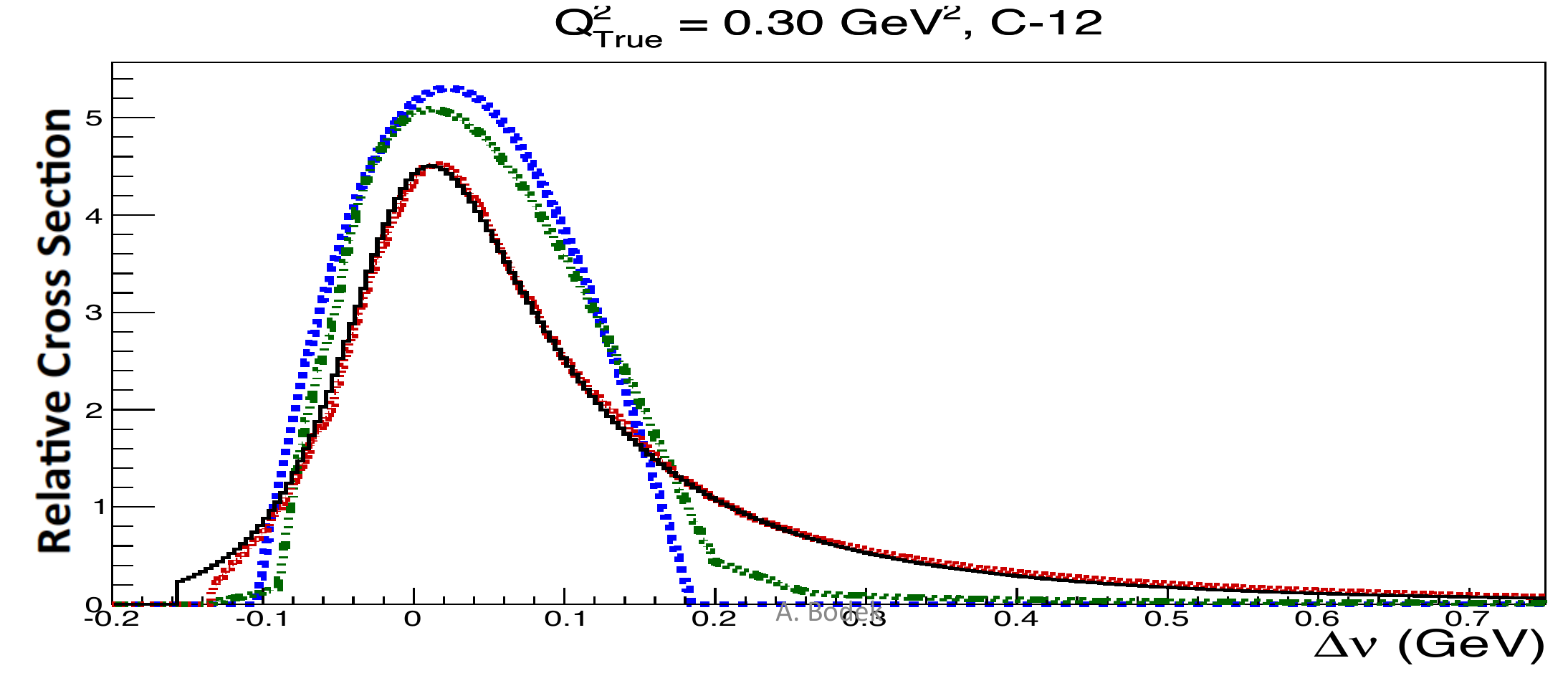}
\includegraphics[width=3.5in,height=2.95in]{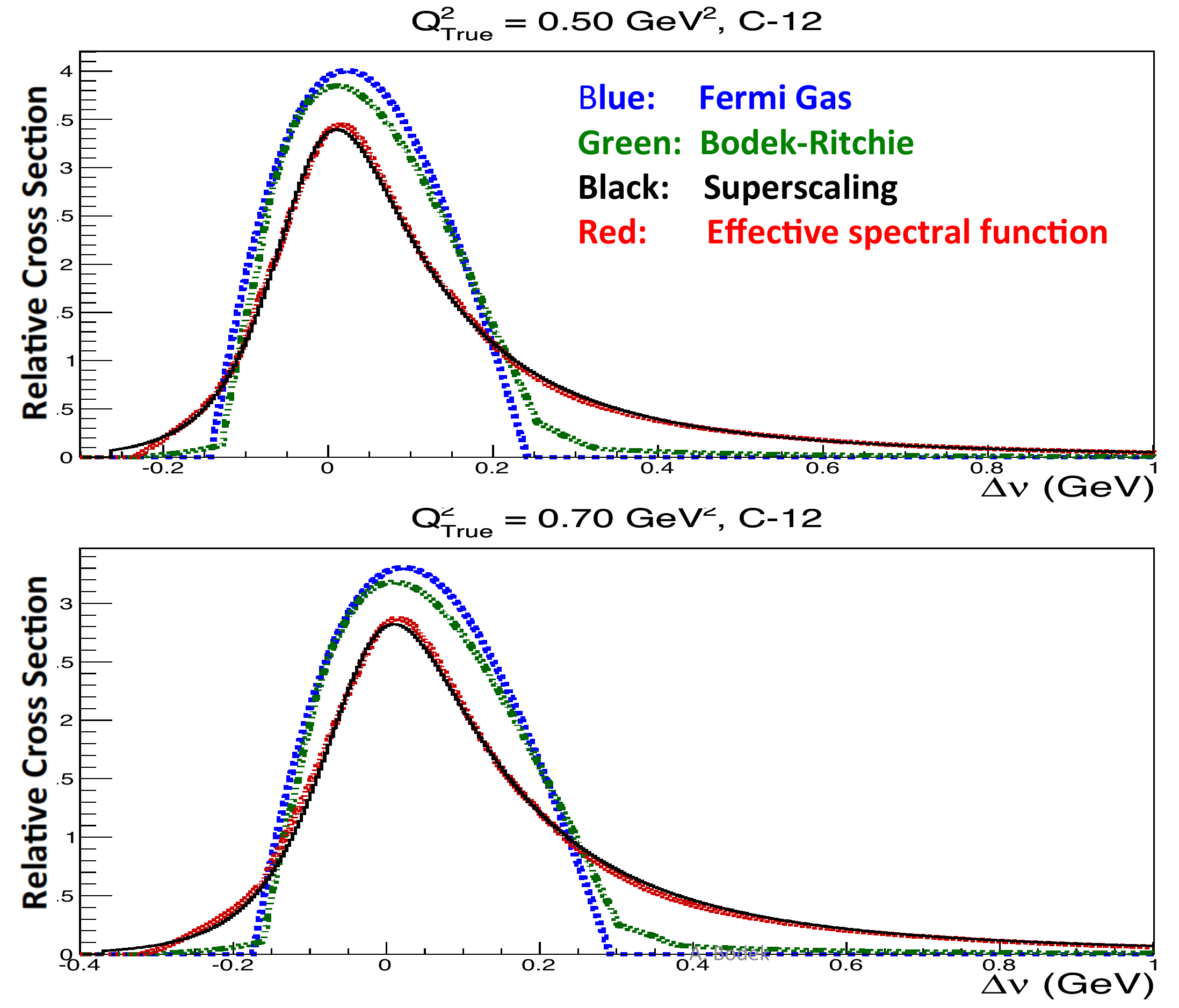}
\includegraphics[width=3.5in,height=2.95in]{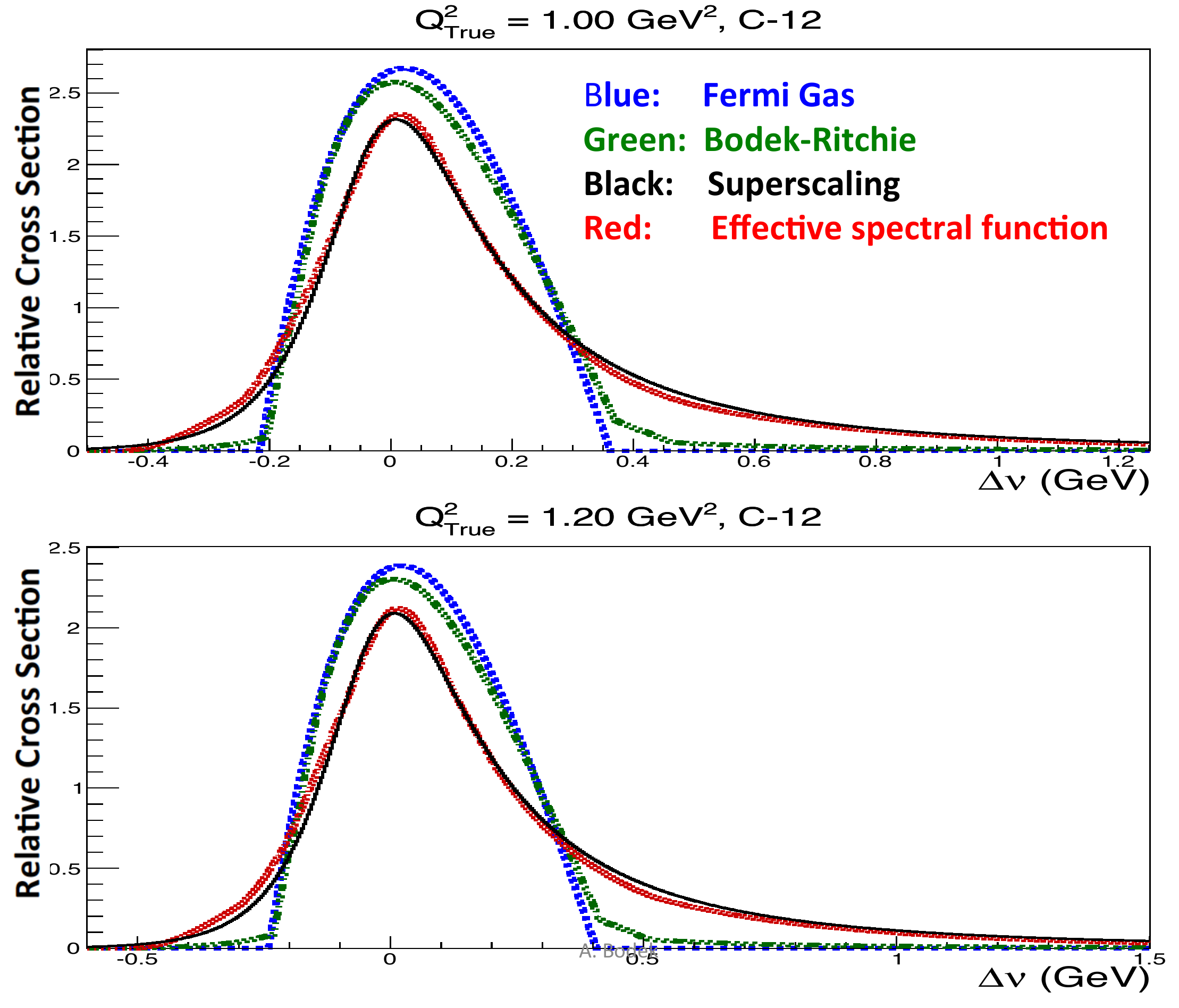}
\caption{ Comparison of the prediction for the normalized QE cross section ($\frac{1}{\sigma} \frac{d\sigma}{d\nu}(Q^2,\nu)$)  for $\carbon$
from the  {\it{effective spectral function}} (red) 
to the predictions of the   $\psi'$ superscaling model (black).  The predictions are shown as
a function of $\Delta\nu$ for  $ Q^2$ values of 0.3, 0.5, 0.7, 1.0, and 1.2 GeV$^2$.   Also shown (for reference) 
are the predictions for the Fermi Gas model in blue, and the predictions from the  Bodek-Ritchie\cite{Bodek-Ritchie} Fermi gas model which includes a high momentum
contribution from two nucleon correlations in green (color online). }
\label{lowhighQ}
\end{center}
\end{figure} 
%
\begin{figure}
\begin{center}
\includegraphics[width=3.5in,height=2.95in]{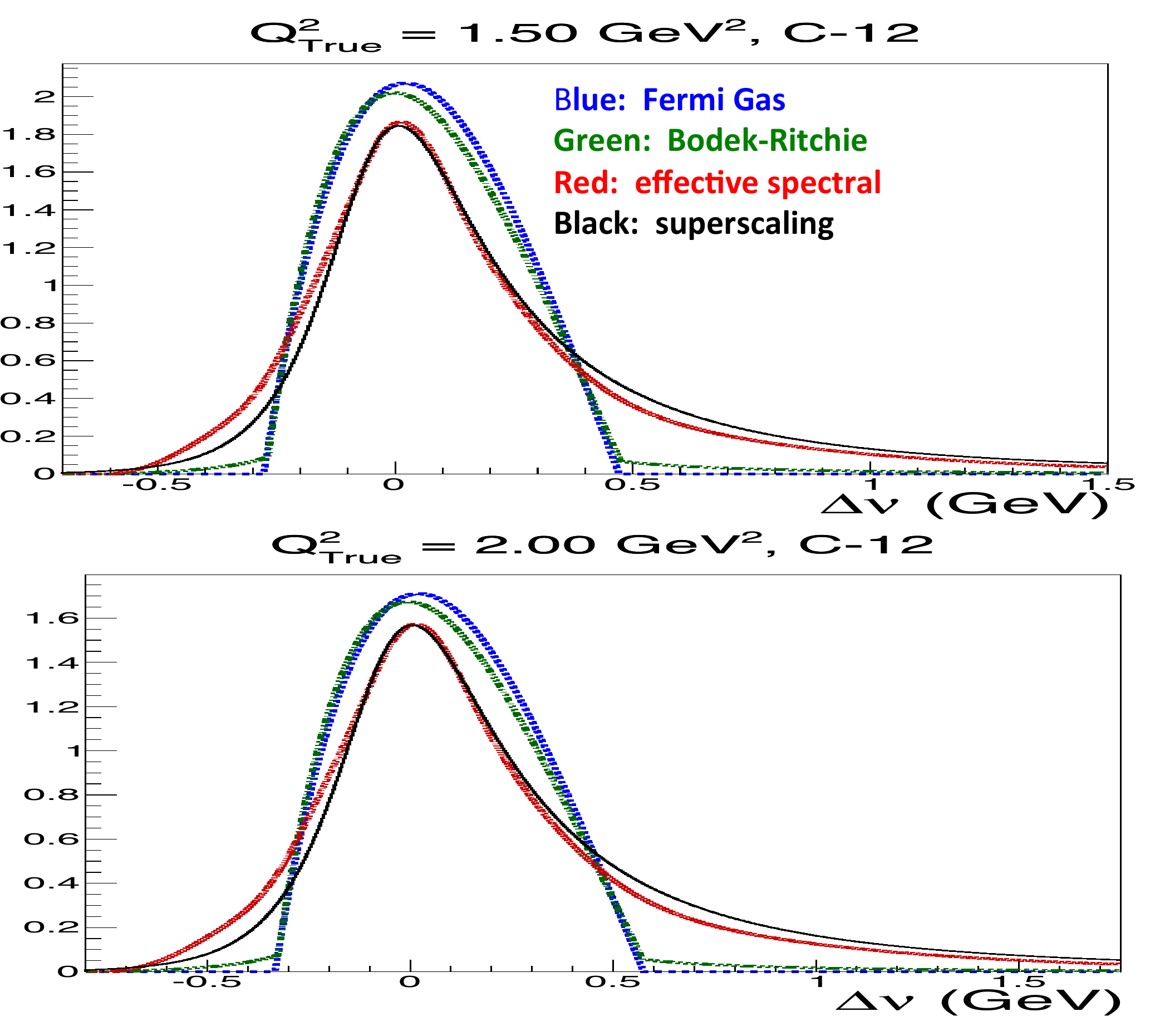}
\caption{ Comparison of the prediction for the normalized QE cross section ($\frac{1}{\sigma} \frac{d\sigma}{d\nu}(Q^2,\nu)$)  for $\carbon$
from the  {\it{effective spectral function}} (red) 
to the predictions of the   $\psi'$ superscaling model (black).  The predictions are shown as
a function of $\Delta\nu$ for  $ Q^2$ values of 1.5 and 2.0  GeV$^2$.   Also shown (for reference) 
are the predictions for the Fermi Gas model in blue, and the predictions from the  Bodek-Ritchie\cite{Bodek-Ritchie} Fermi gas model which includes a high momentum
contribution from two nucleon correlations in green (color online). }
\label{highQ}
\end{center}
\end{figure} 
%
%
\begin{figure}
\begin{center}
\includegraphics[width=3.5in,height=3.0in]{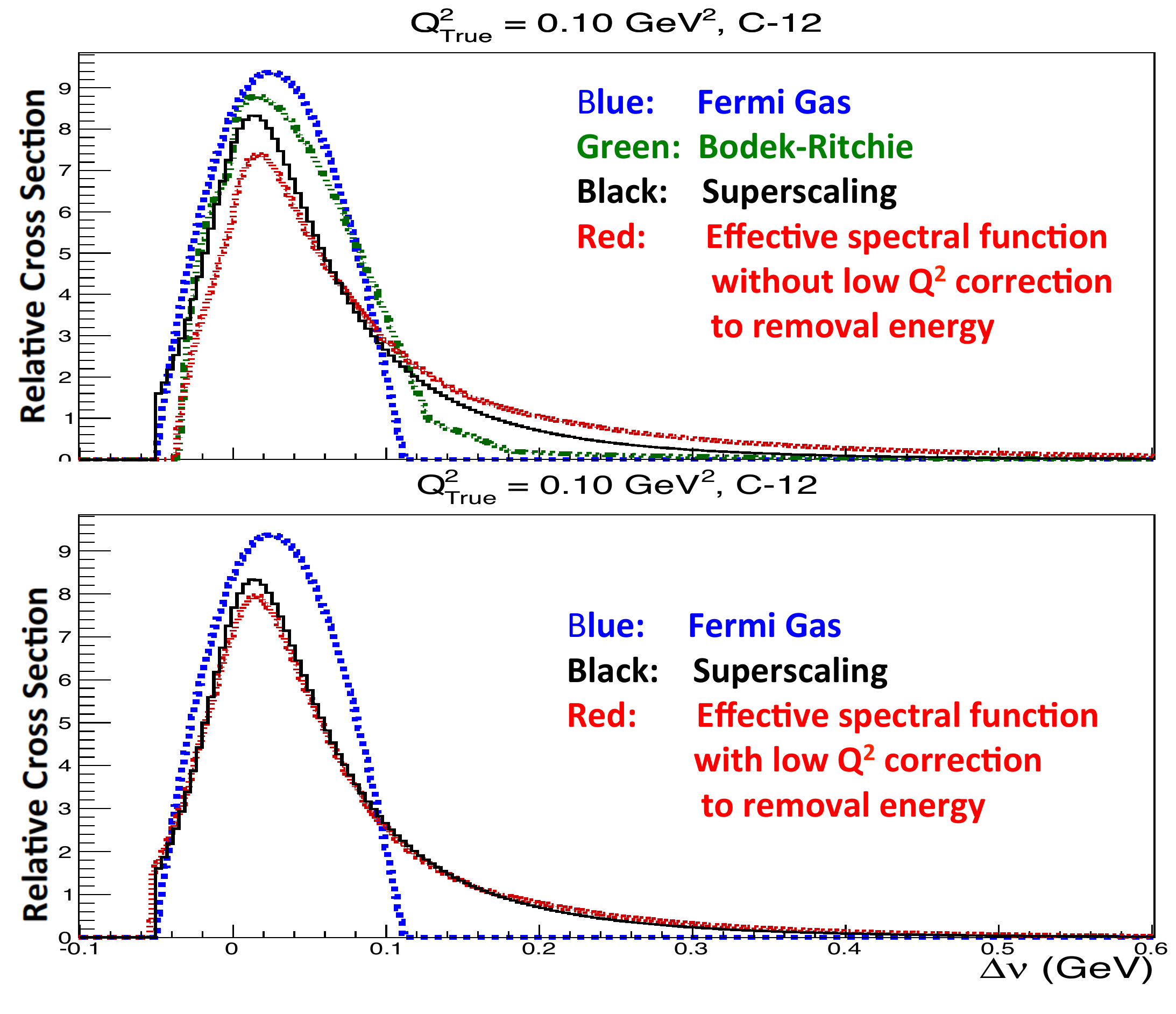}
\caption{ Comparison of the prediction for the normalized QE cross section ($\frac{1}{\sigma} \frac{d\sigma}{d\nu}(Q^2,\nu)$)  for $\carbon$
from the  {\it{effective spectral function}} (red) 
to the predictions of the   $\psi'$ superscaling model (black).  The predictions are shown as
a function of $\Delta\nu$ for  $ Q^2$ = 0.1 GeV$^2$.   The top panel show the predictions without the low $Q^2$ correction factor to the removal energy,
The bottom panel shows the predictions including the  low $Q^2$ correction factor to the removal energy
e.g. from  final state interaction (of the first kind)  at low $Q^2$. Also shown (for reference) 
is the prediction for superscaling in black,
 the predictions for the Fermi Gas model in blue, and the predictions from the  Bodek-Ritchie\cite{Bodek-Ritchie} Fermi gas model which includes a high momentum
contribution from two nucleon correlations in green (color online). }
\label{lowQ}
\end{center}
\end{figure} 
%
\begin{figure}
\begin{center}
\includegraphics[width=3.5in,height=3.5in]{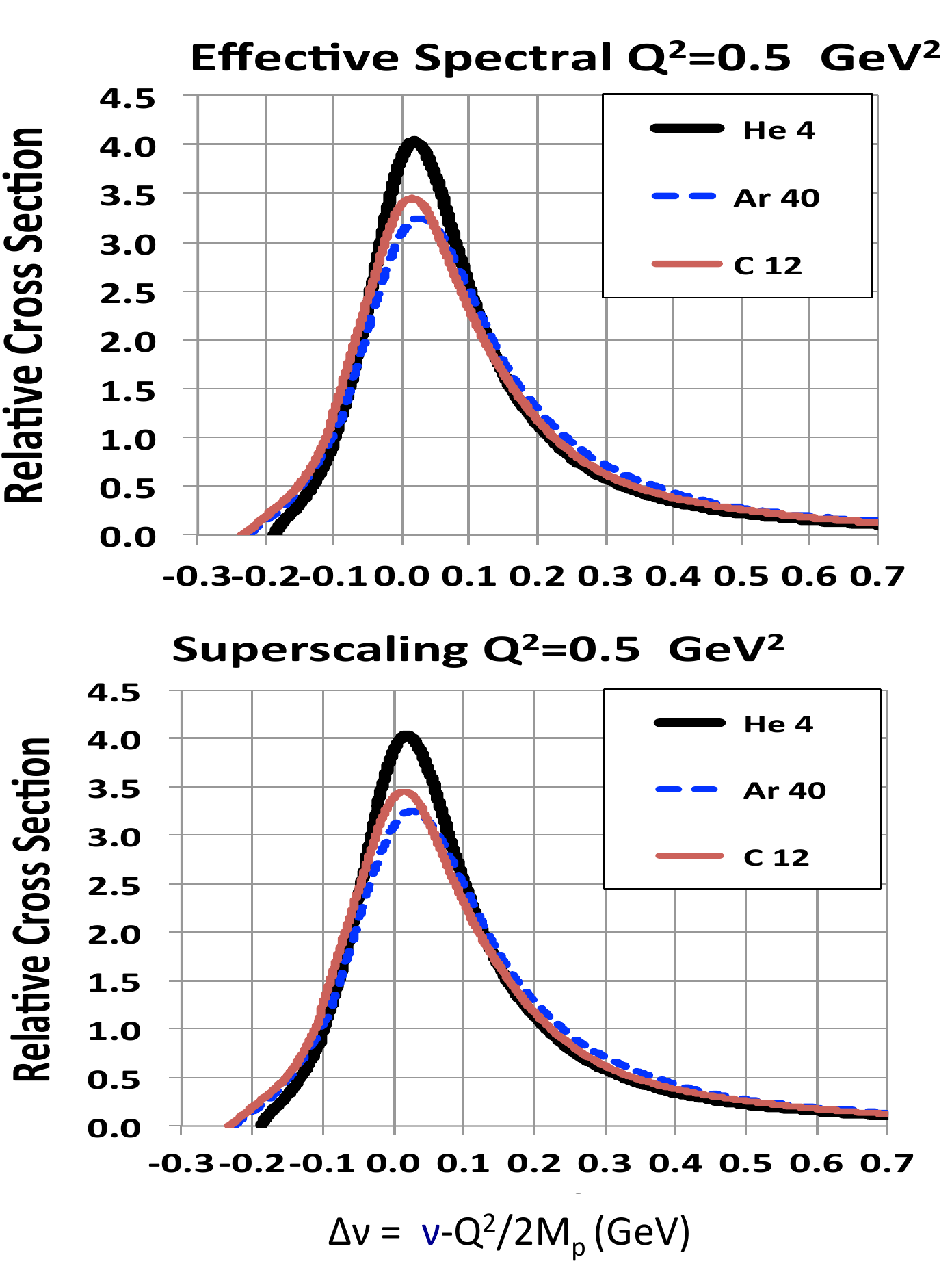}
\caption{Predictions  for the  normalized QE cross section ($\frac{1}{\sigma} \frac{d\sigma}{d\nu}(Q^2,\nu)$) 
 from the  {\it{effective spectral function} } (top panel) as compared to the
predictions of  $\psi'$  superscaling (bottom panel) for different nuclei.  The predictions are shown as
a function of $\Delta\nu$ for $Q^2$=0.5 GeV$^2$ (color online).}
\label{nuclei}
\end{center}
\end{figure} 
%
%
\begin{figure}
\begin{center}
\includegraphics[width=3.5in,height=2.in]{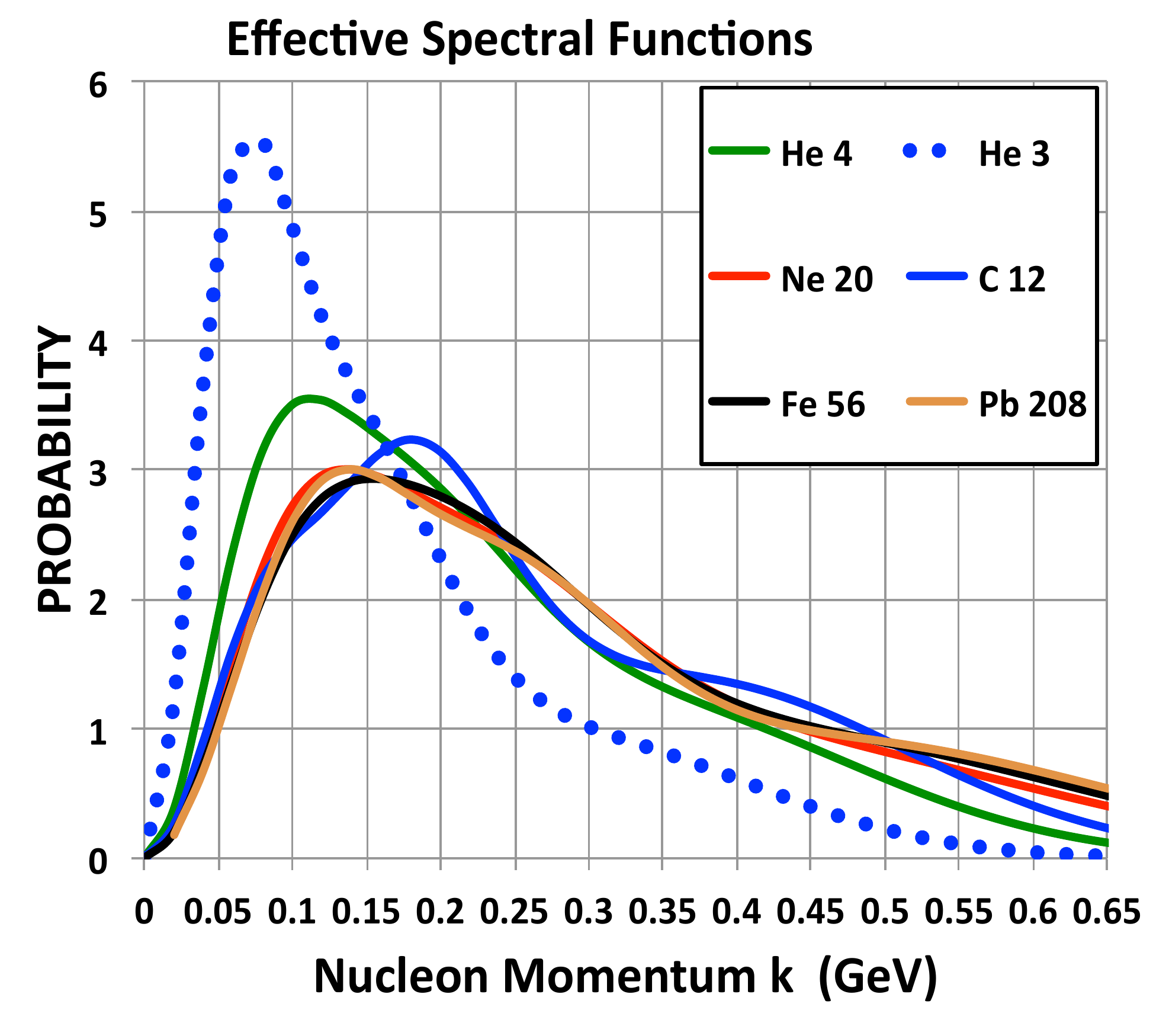}
\caption{The  momentum distributions for the  {\it{effective spectral function}} for various nuclei
($\rm ^{3}He$,  $\rm ^{4}He$,  $\rm ^{12}C$  $\rm ^{20}Ne$,   $\rm ^{56}Fe$, and $\rm ^{208}Pb$). In the analysis we
set all distributions to zero for $k>$0.65 GeV (color online).}
\label{momentumA}
\end{center}
\end{figure} 
\subsubsection{Comparison of the 1p1h and 2p2h contributions}
The top panel of Fig. \ref{1p1h2p2h} shows  the prediction of the  {\it{effective spectral function}} model 
for  $\frac{1}{\sigma} \frac{d\sigma}{d\nu}(Q^2,\nu)$ for QE scattering from a 
   $\carbon$ nucleus  at $Q^2$ = 0.5 GeV$^2$,  assuming that only the 1p1h process contributes.
    The  bottom panel of Fig. \ref{1p1h2p2h} shows  the prediction of the  {\it{effective spectral function}} model 
for  $\frac{1}{\sigma} \frac{d\sigma}{d\nu}(Q^2,\nu)$ for QE scattering from a 
   $\carbon$ nucleus  at $Q^2$ = 0.5 GeV$^2$ assuming that only the 2p2h process contributes.

 We find that the  {\it{effective spectral function}}  
with only the 1p1h process provides a reasonable description of the prediction of  $\psi'$ superscaling.  Including a  contribution from the 2p2h process in the fit  improves the agreement and results in  a prediction which is almost identical to the prediction of  $\psi'$ superscaling.

For reference, figures  \ref{1p1h2p2h}- \ref{lowQ} also show 
 the prediction for the Fermi Gas model in blue, and the predictions from the  Bodek-Ritchie\cite{Bodek-Ritchie} Fermi gas model which includes a high momentum
contribution from two nucleon correlations in green. These predictions are calculated for 10 GeV neutrinos
using the  the GENIE neutrino Monte Carlo generator.

%
\subsubsection{Comparisons as a function of $Q^2$ for $Q^2>0.3$ GeV$^2$}
\label{sectionfit}

Fig. \ref{lowhighQ}   and  Fig. \ref{highQ} shows a  comparison of the prediction  of the  {\it{effective spectral function}} for  $\frac{1}{\sigma} \frac{d\sigma}{d\nu}(Q^2,\nu)$ for $\carbon$  to the predictions of the  $\psi'$ superscaling formalism 
 for $Q^2$ values of 0.3, 0.5, 0.7, 1.0, 1.2, 1.5 and 2.0   and GeV$^2$. 

 In principle, it should not be possible for a spectral function approach
to exactly reproduce  $\psi'$ superscaling at all values of $Q^2$.  Nonetheless,   the parameters which we optimized for $Q^2$  values of
0.3, 0.5 and 0.7  GeV$^2$ also provide a good  description of  $\frac{1}{\sigma} \frac{d\sigma}{d\nu} (\nu)$ for 
$Q^2$ values of 1.0, 1.2, 1.5 and 2.0 GeV$^2$.
 

\subsubsection{Comparisons as a function of $Q^2$ for $Q^2<0.3$ GeV$^2$ }

The  low $Q^2$ suppression factor of the removal energy which is 
 given in  eq. \ref{Veq} is introduced in 
  order to  reproduce  predictions of the  $\psi'$ superscaling model at $Q^2<0.3$ GeV$^2$.
 
   Fig. \ref{lowQ}  shows a  comparison of the prediction for the shape ($\frac{1}{\sigma} \frac{d\sigma}{d\nu}(Q^2,\nu)$) of the quasielastic  peak for $\carbon$
from the {\it{effective spectral function}} (red) 
to the predictions of the   $\psi'$ superscaling model (black).  The predictions are shown as
a function of $\Delta\nu$ for  $ Q^2$ = 0.1 GeV$^2$. 
The top panel show the predictions without the low $Q^2$ correction factor to the removal energy.
The bottom panel shows the predictions including the  low $Q^2$ correction factor to the removal energy, e.g. from  final state interaction  (of the first kind) at low $Q^2$. 

Also shown (for reference) 
is the prediction for superscaling  in black, the prediction for the Fermi Gas model in blue, and the predictions from the  Bodek-Ritchie\cite{Bodek-Ritchie} Fermi gas model which includes a high momentum
contribution from two nucleon correlations in green.
%
\section{Spectral functions for other  nuclei }
Fig. \ref{nuclei} shows the predictions for the  shape of the quasielastic  peak  ($\frac{1}{\sigma} \frac{d\sigma}{d\nu} (\nu)$) from the best fit 
 {\it{effective spectral function} } (top panel) as compared to the
predictions of  $\psi'$  superscaling model (bottom panel) for different nuclei. The predictions are shown as
a function of $\Delta\nu$ for $Q^2$=0.5 GeV$^2.$
Fig. \ref{momentumA} shows the momentum distribution of the {\it{effective spectral function}}  for various nuclei, and
Table \ref{fitsA} gives the  parameterizations of the  {\it{effective spectral function}}  for various nuclei.
%
%
\begin{table*}
\begin{center}
\begin{tabular}{|c|c|c|c|c|c|c|c|c|} \hline 
Parameter  & $\rm ^{3}He$    & $\rm ^{4}He$    &  $\rm ^{12}C$   & $\rm ^{20}Ne$   & $\rm ^{27}Al$  &  $\rm ^{40}Ar$ &  $\rm ^{56}Fe$
 &  $\rm ^{208}Pb$ \\ \hline\hline
$\Delta$(MeV)   &   5.3      &   14.0    &  12.5      & 16.6      & 12.5    & 20.6     & 15.1        & 18.8   \\ \hline
$f_{1p1h}$&  0.312    &  0.791   &    0.808  & 0.765    & 0.774  & 0.809    & 0.822     & 0.896  \\ \hline
$b_s$       &  3.06     &    2.14   &   2.12    & 1.82    & 1.73     & 1.67      & 1.79       & 1.52  \\ \hline
$b_p$       &  0.902   &   0.775  &  0.7366   & 0.610   & 0.621 & 0.615  & 0.597    & 0.585  \\ \hline
$\alpha$    &  10.93    & 9.73   &   12.94   & 6.81 & 7.20 & 8.54 & 7.10 & 11.24   \\ \hline
$\beta$      &  6.03      &   7.57  &  10.62        & 6.08 & 6.73  & 8.62    &  6.26     & 13.33 \\ \hline
$c_1$        &   199.6    &  183.4 &  197.0       & 25.9  & 21.0   & 200.0  & 18.37 & 174.4   \\ \hline
$c_2$       & 1.92      & 5.53  &  9.94             & 0.59 & 0.59   & 6.25 & 0.505     & 5.29  \\ \hline
$c_3$         & 5.26x10$^{-5}$     &  59.0x10$^{-5}$ & 4.36 x10$^{-5}$      & 221. x10$^{-5}$   & 121.5 x10$^{-5}$  & 269.0x10$^{-5}$ & 141.0 x10$^{-5}$ & 9.28x10$^{-5}$   \\\hline
$N$         &  6.1   & 18.94   &  29.64         & 4.507  & 4.065   & 40.1  & 3.645  & 37.96     \\ \hline\end{tabular}
\caption{ Parameterizations of the  {\it{effective spectral function}}  for various nuclei. 
Here, $\Delta$ is the binding energy parameter, and
$f_{1p1h}$ is the fraction of the scattering that occurs via the ${1p1h}$ process.  For deuterium ($\deuteron$) are see Table \ref{fitsC}.}
\label{fitsA}
\end{center}
\end{table*}
%
%
\begin{figure}
\begin{center}
\includegraphics[width=3.5in,height=2.in]{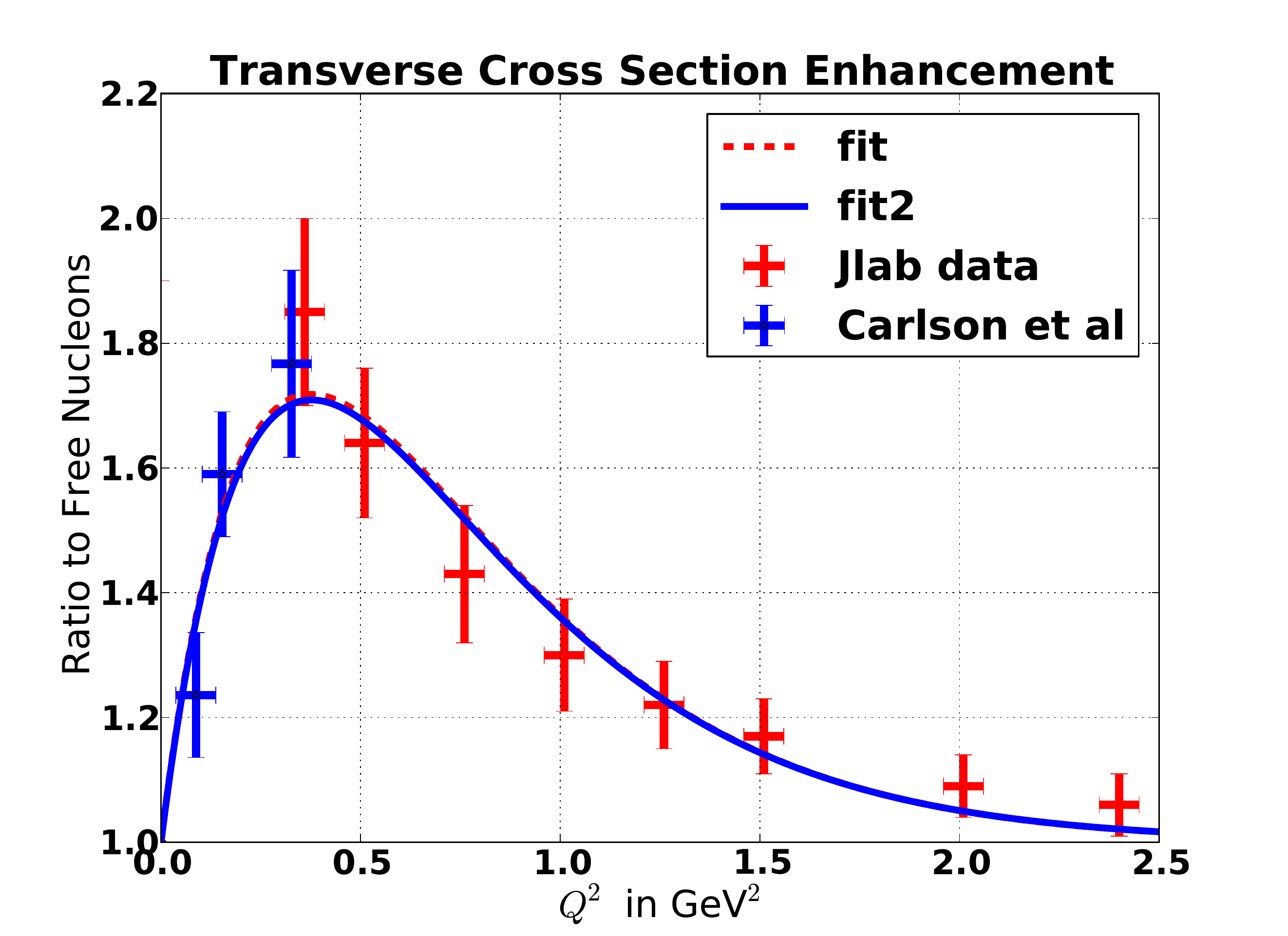}
\includegraphics[width=3.5in,height=2.in]{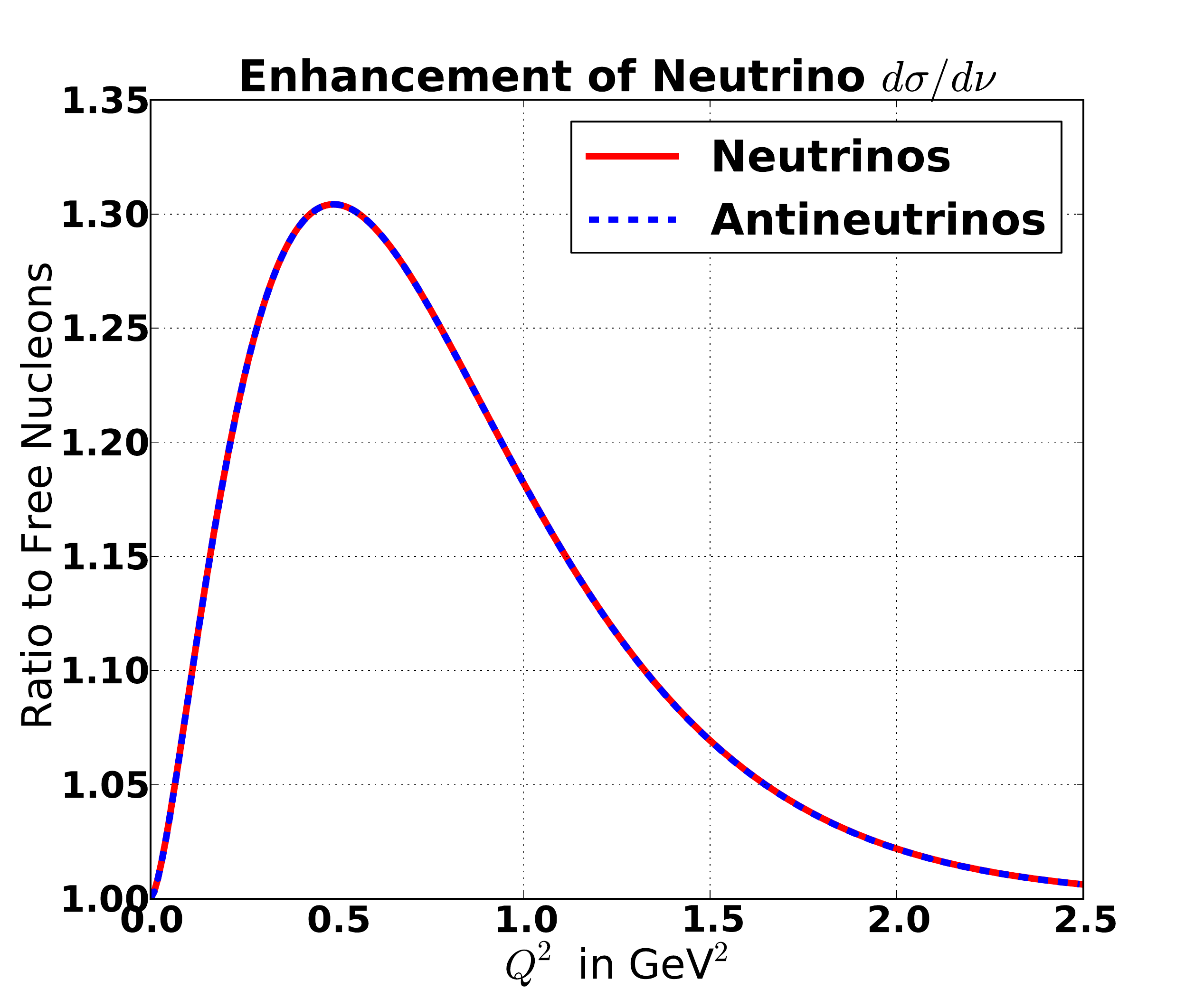}
\caption{Top panel:   Electron scattering data. The ratio of the transverse  QE cross section with TE to the transverse cross section for free nucleons
as a function of $Q^2$. Bottom panel:  
The ratio of $d\sigma/dQ^2$  neutrino QE  cross ion carbon (with TE)  to the  sum of free nucleon cross sections as a function
of  $Q^2$ for neutrino and antineutrino energies above 3 GeV.}
\label{GMPN}
\end{center}
\end{figure} 
%
\begin{figure}
\begin{center}
\includegraphics[width=3.5in,height=2.in]{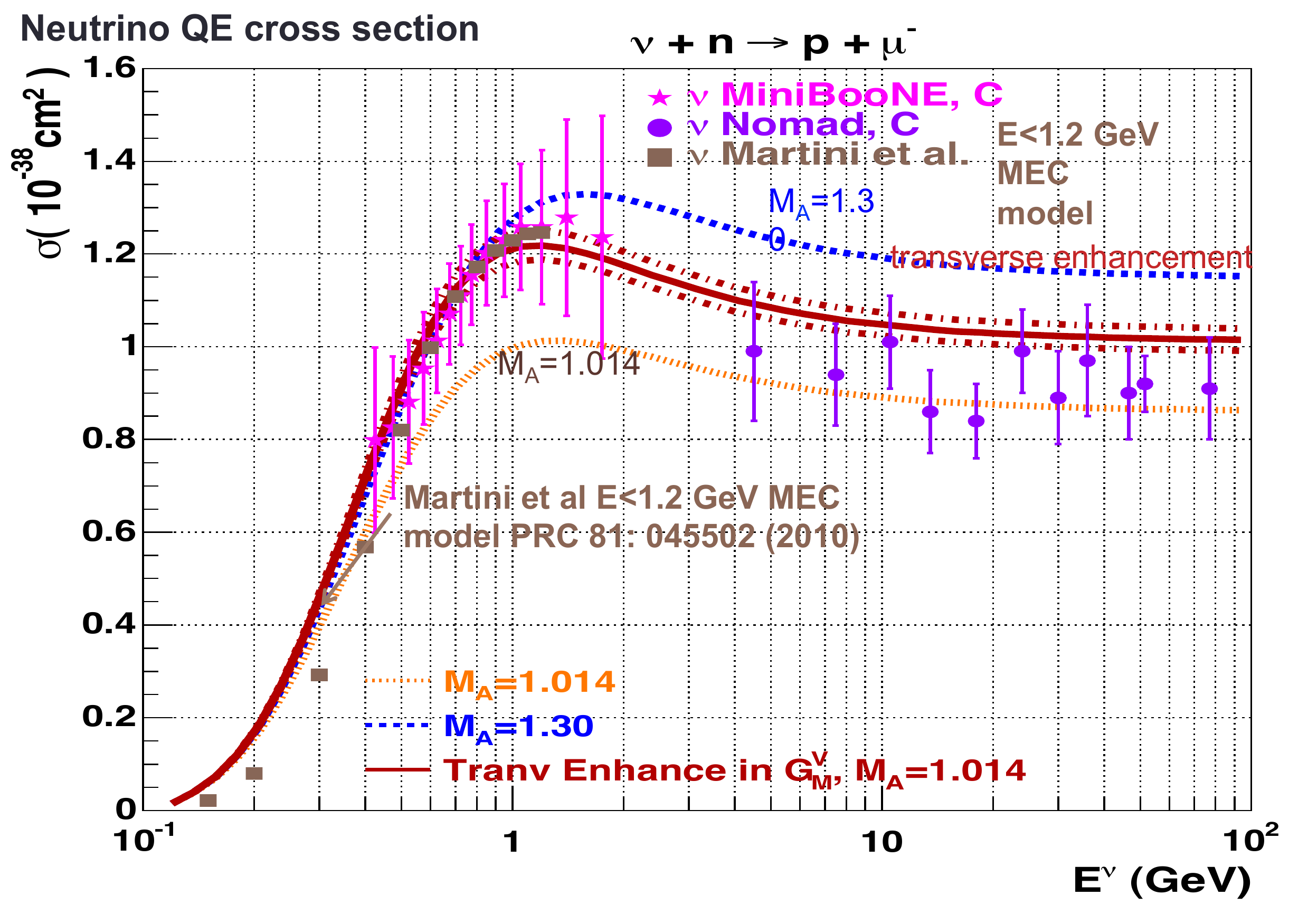}
\includegraphics[width=3.5in,height=2.in]{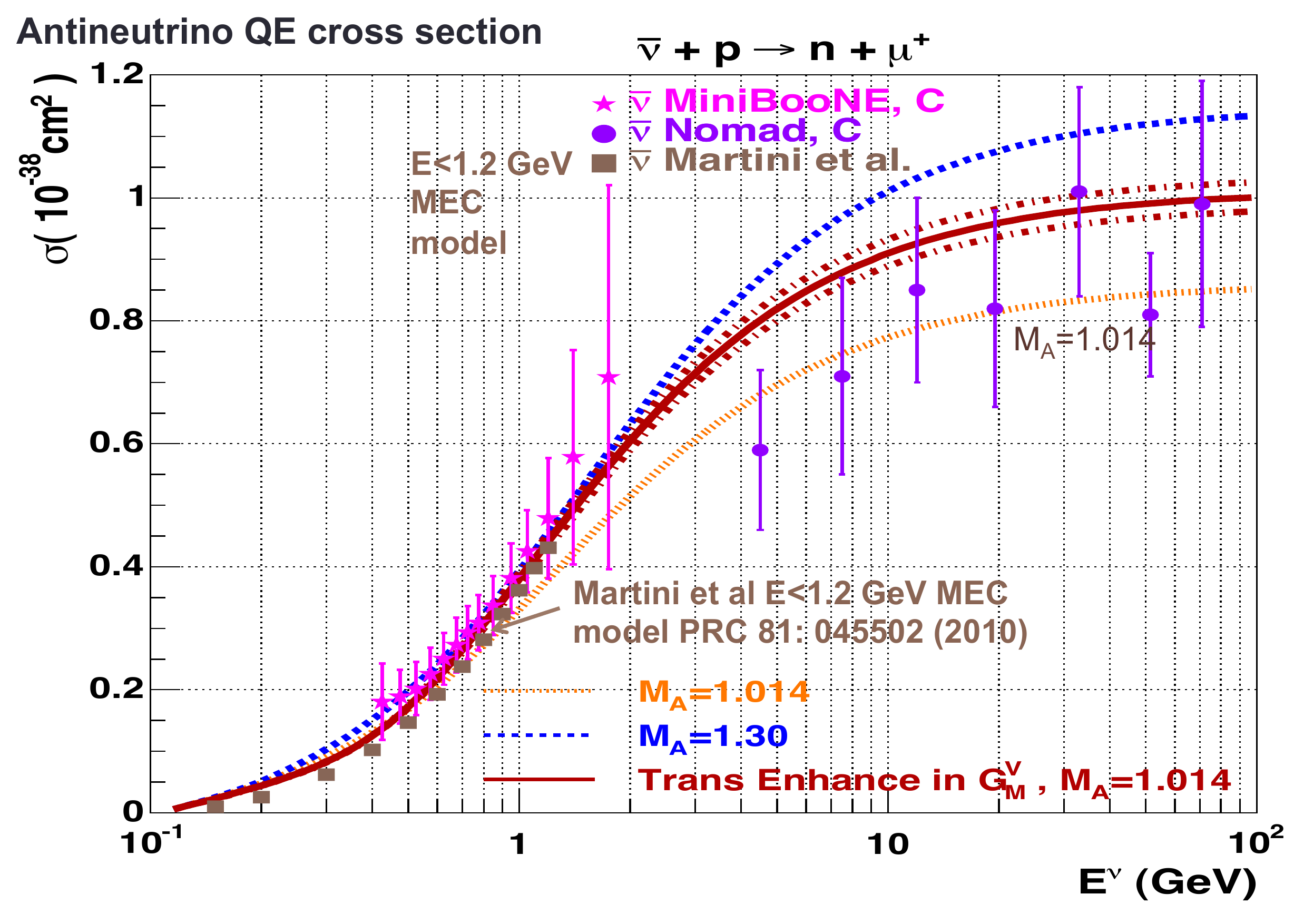}
\caption{
The   neutrino QE  cross section on carbon with  TE  and without TE as a function of neutrino energy.   The cross section for neutrinos is  shown on the top panel and  the cross section for  antineutrinos  is shown in bottom panel. }
\label{nu-cross}
\end{center}
\end{figure} 
%
\begin{figure}
\begin{center}
\includegraphics[width=3.5in,height=2.in]{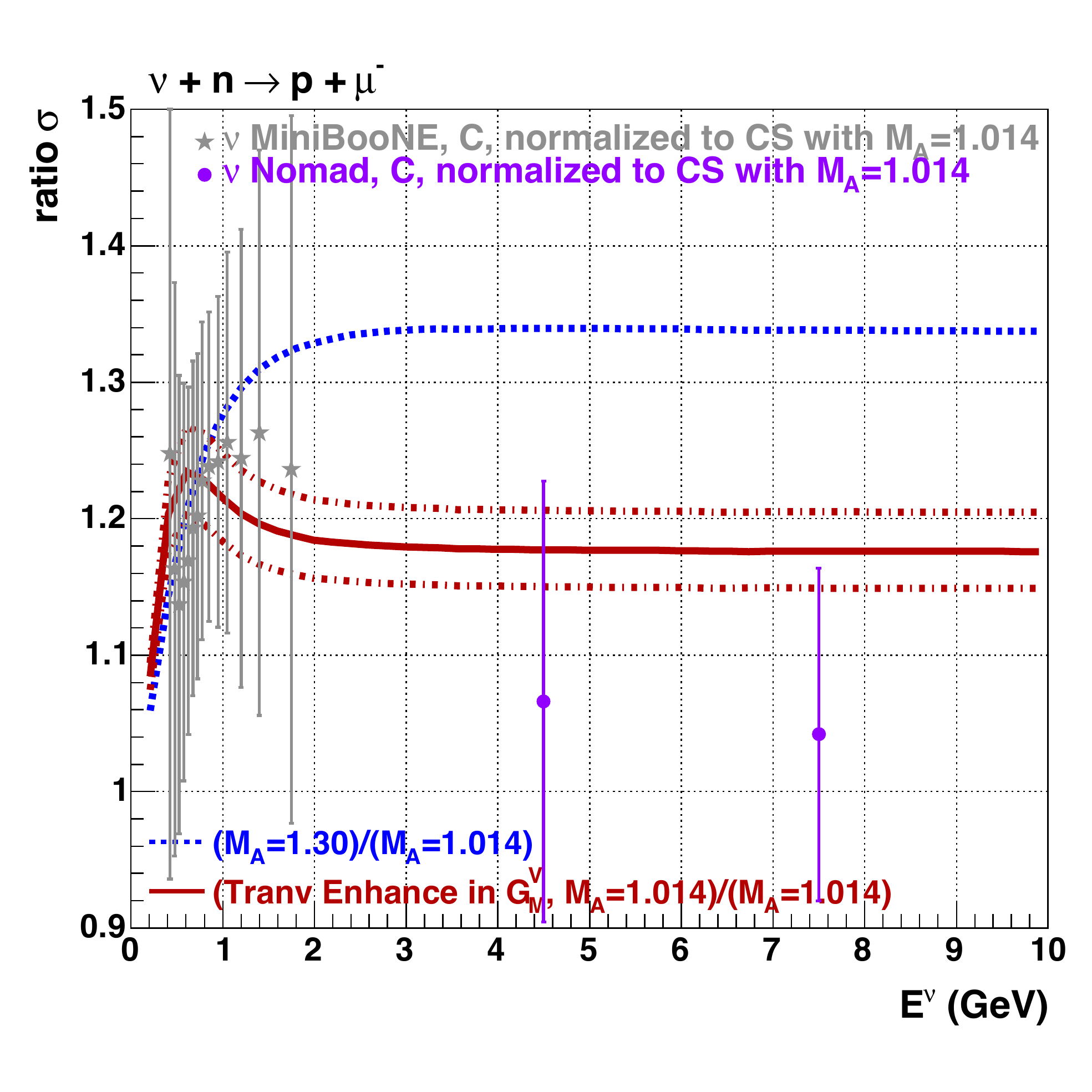}
\includegraphics[width=3.5in,height=2.in]{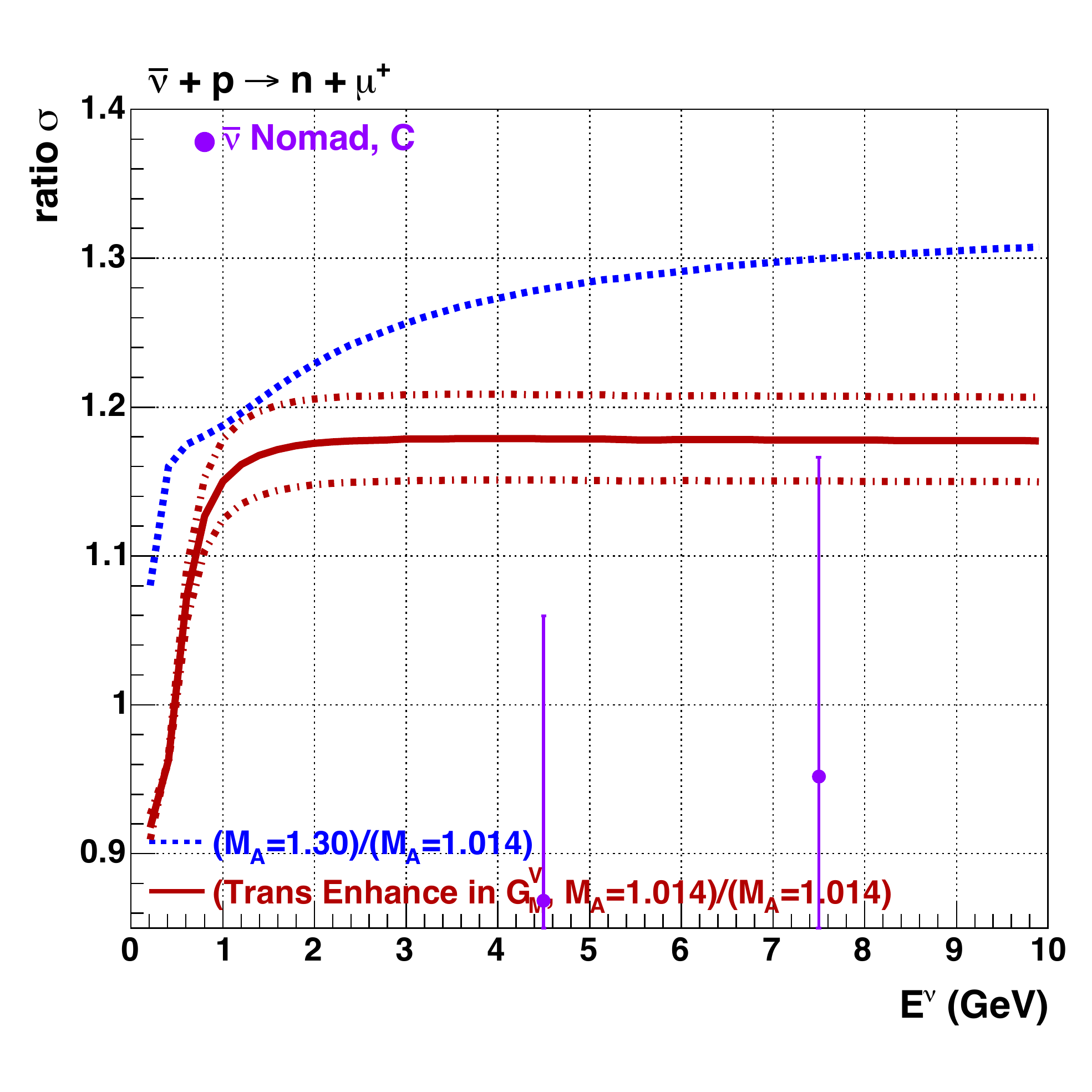}
\caption{
The ratio of the total neutrino QE  cross section on carbon with  TE to sum of free nucleon cross sections  as a function of energy.   The ratio for neutrinos is  shown on the top panel and   the ratio for antineutrinos  is shown in bottom panel. On average the overall cross section  is increased  by  about 18\%. }
\label{TE_energy}
\end{center}
\end{figure} 
\begin{figure}
\includegraphics[width=3.5in,height=2.5in]{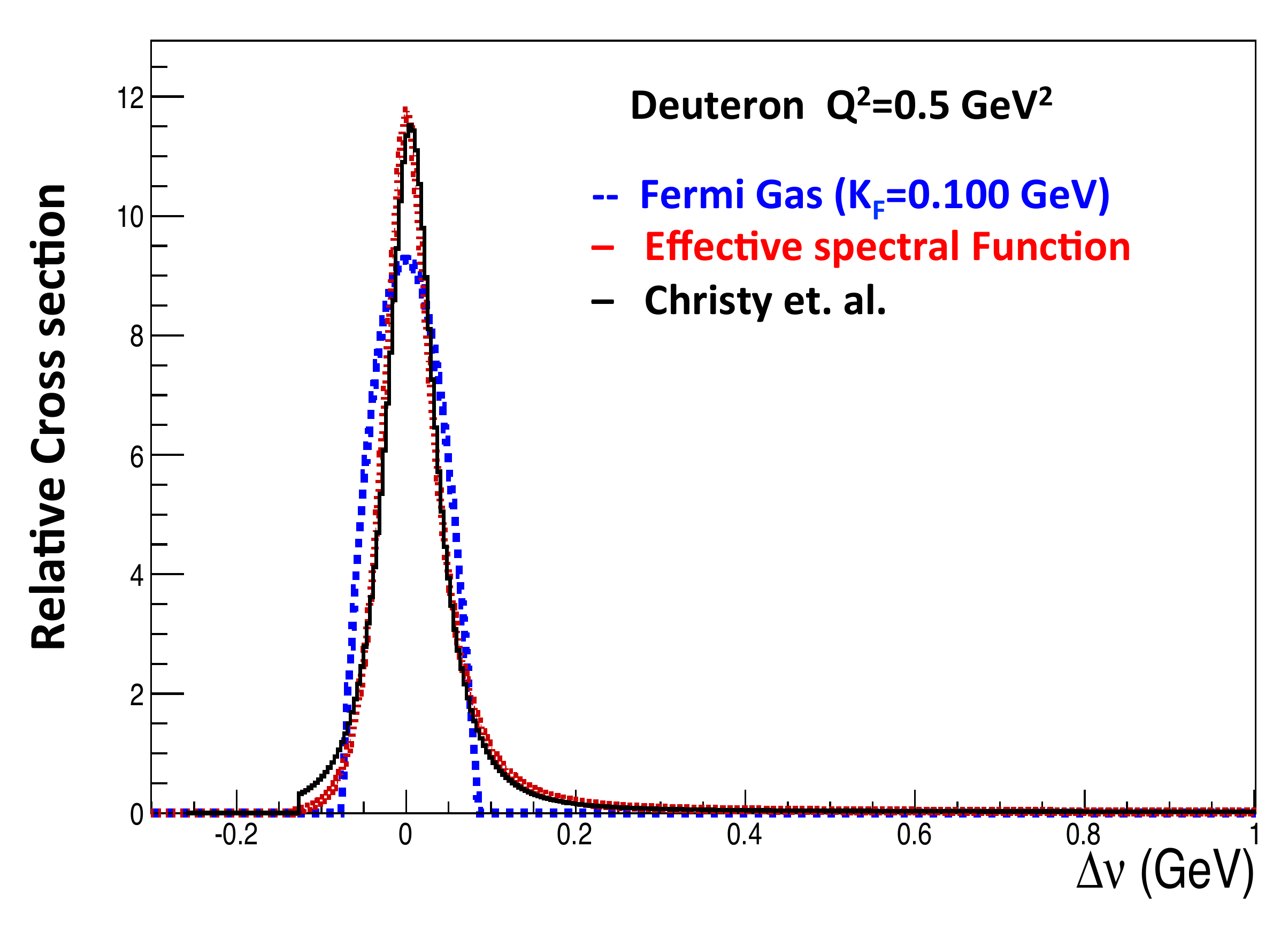}
\caption{ Comparisons of model predictions for   $\frac{1}{\sigma} \frac{d\sigma}{d\nu}(Q^2,\nu)$  as a function
of $\Delta \nu$ = $\nu -Q^2/2M$  for  QE  electron scattering on the deuterium  
at $Q^2$=0.5 GeV$^2$.  The solid black line is the prediction from reference\cite{D2} (which agrees with electron scattering
data). The red line is the prediction of the best fit parameters for  {\it{effective spectral function}}  the deuteron.  The blue line is the prediction
for a Fermi gas with a Fermi momentum $K_F$ = 0.100 GeV.  The  predictions with the {\it{effective spectral function}} 
are in agreement with the calculations of reference\cite{D2} (color online).}
\label{D-0.5}
\end{figure} 
%
\begin{figure}
\begin{center}
\includegraphics[width=3.5in,height=2.3in]{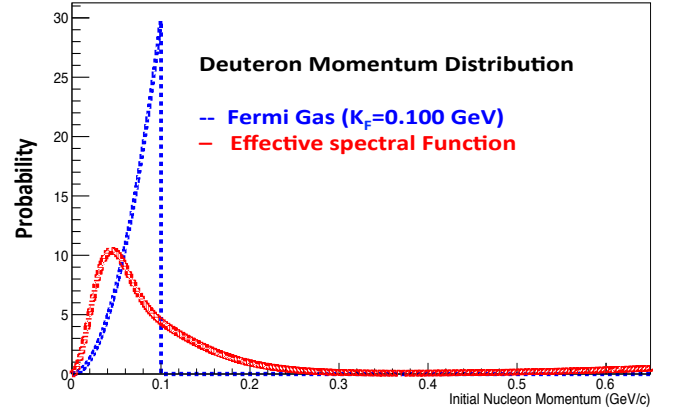}
\caption{The  momentum distribution for the  {\it{effective spectral function}}  for deuterium (shown in red). Also
shown (in blue)  is the momentum distribution for a Fermi gas with $K_F$=0.10 GeV. Note that we set the probability to
zero for $k>K_M$ where  $K_M$=0.65 GeV (color online).}
\label{momentumD}
\end{center}
\end{figure} 
%
%
\begin{figure}
\begin{center}
\includegraphics[width=3.5in,height=2.3in]{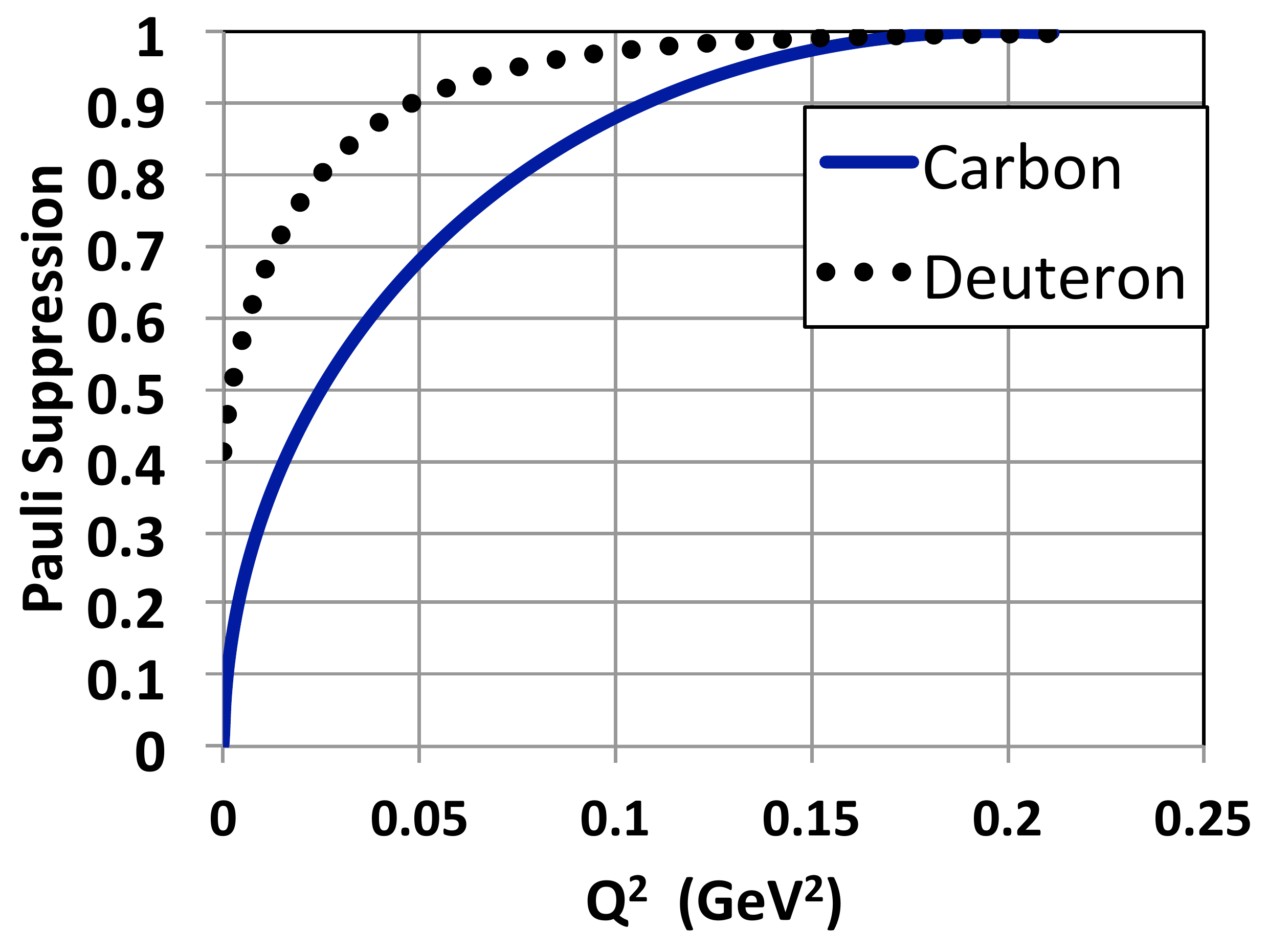}
\caption{A comparison of the Pauli suppression factor $K_{Pauli}$ for carbon and the deuteron as a function
of the square of the four momentum transfers $Q^2$.  For the carbon factor (eq. \ref{nucpauli}) we assumed 
$|\bf {q} |$ =$\sqrt{Q^2(1+Q^2/(4M_p^2))}$  (color online).}
\label{DCpauli}
\end{center}
\end{figure} %
\section{Transverse Enhancement and multi nucleon processes}
Both spectral functions and  the $\psi'$ superscaling formalism 
model  QE scattering in terms of scattering from independent nucleons in the initial state.
The independent nucleon approach works well in modeling the
longitudinal (electric) cross section for QE electron scattering
from nuclear targets.
However, it is known that
none of the independent nucleon models can describe the
transverse (magnetic)  part of the QE electron scattering cross section.  The
transverse cross section is larger than the predictions of 
the independent nucleon model, and the enhancement is
a function of $Q^2$. This experimental observation is
usually  referred to as "Transverse Enhancement" (TE).

TE has been attributed to multi nucleon processes such as meson
exchange currents and isobar excitation. There are two
ways to account for these multi nucleon effects.

\subsection{Accounting for  contributions from multi nucleon processes}
One way to account for TE is to use specific theoretical
models to estimate the 
contributions from various multi nucleon processes such as meson
exchange currents and isobar excitation.  The differences between
various model may provide an indication of the uncertainties of
the calculations.

\subsection{The Transverse Enhancement (TE) model}
Another way to account for  transverse enhancement  is to parameterize
the experimentally observed excess in  the electron scattering data in a model
independent way. In the  TE model \cite{TE} 
this is done by modifying the magnetic form factors
for bound nucleons. 

As was done in reference \cite{TE}, we have re-extracted the
 integrated transverse enhancement ratio  ${\cal R}_{T}$,
where  $${\cal R}_{T} =\frac {(QE_{transverse}+TE)_{exp}}{(QE_{transverse})_{model}},$$from electron scattering data
from the JUPITER collaboration\cite{JUPITER}.   Here  $ {(QE_{transverse}+TE)_{exp}}$ is the experimentally
observed integrated transverse QE cross section  and  $(QE_{transverse})_{model}$ is the integrated  transverse QE  cross section predicted by $\psi'$ super scaling.

 Fig.~\ref{GMPN} shows experimental  values of ${\cal R}_{T}$  as a function of $Q^2$.  The black points are extracted from Carlson $et~al$\cite{MEC4}, and
the higher $Q^2$ are re-extracted  from QE data from the JUPITER collaboration\cite{JUPITER}.   The  $Q^2$ dependence of ${\cal R}_{T}$  is parametrized by the expression:   $${\cal R}_{T}=1+AQ^2e^{-Q^2/B}$$    with A=5.194 and B=0.376~GeV$^2$. 
    The electron scattering data indicate that the transverse enhancement is maximal near $Q^2$=0.3~GeV$^2$ and is small for $Q^2$ greater
than 1.5~GeV$^2$ This parametrization     is valid for carbon (A=12)  (it is also  an approximate representation for higher A  nuclei).

We assume that the enhancement in the transverse QE cross 
section can be described by a modification of  ${\cal G}_M^V(Q^2) = G_{Mp}(Q^2)-G_{Mn}(Q^2)$ for nucleons bound in carbon,
where $G_{Mp}$ and $G_{Mn}$ are the magnetic form factor of the proton and neutron, respectively (as measured in 
electron scattering experiments).
We use the parametrization of ${\cal R}_{T}$ to modify $G_{Mp}$ and $G_{Mn}$ for bound 
nucleons as follows.  
 \begin{eqnarray}
  {G_{Mp}^{nuclear}(Q^2)}&=&  G_{Mp}(Q^2) \times \sqrt {1+AQ^2e^{-Q^2/B}} 
     \nonumber \\
    {G_{Mn}^{nuclear}(Q^2)} &=&  G_{Mn}(Q^2) \times  \sqrt {1+AQ^2e^{-Q^2/B}}. 
 \label{magnetic} 
      \end{eqnarray}   

 Transverse enhancement increases the overall neutrino and antineutrino cross sections
 and changes the shape of the  differential cross
section as a function of $Q^2$ as shown in the bottom panel of  Fig. \ref{GMPN}.

Fig.  \ref{nu-cross} shows the   neutrino and antineutrino QE  cross sections on $\carbon$ with  TE  and without TE as a function of neutrino energy.   The cross section for neutrinos is  shown on the top panel and  the cross section for  antineutrinos  is shown in bottom panel. 
 Fig. \ref{TE_energy} shows the ratio of the  neutrino and antineutrino QE  cross sections on $\carbon$ with  TE to the sum of free nucleon cross sections  as a function of energy.   The ratio for neutrinos is  shown in the top panel and  the ratio for antineutrinos  is shown inn the  bottom panel. On average the overall cross section  is increased  by  about 18\%.   

 Note that TE is a 2p2h process. Therefore, 
when TE is included in the model prediction, the relative fractions of
the 1p1h and 2p2h should be changed as follows:
%
 \begin{eqnarray}
f_{1p1h}^{(with~TE) }&= &\frac{ f_{1p1h}}{1.18} \nonumber \\
f_{2p2h}^{(with~TE)} &=&\frac{f_{2p2h}+0.18}{1.18}
\label{eq-frac}
\end{eqnarray}

In the above prescription, the 
energy sharing between the two nucleons in the final
state for the 2p2h TE process is the same as for the 2p2h process from short
range two nucleon correlations.  We can make other assumptions 
about  the  energy sharing between the two nucleus for the TE process.  For example
one can chose to use a  uniform angular
distribution of the two nucleons in the center of mass of the two nucleons as is done in NuWro\cite{nuwro}.  This
can easily be done in a neutrino MC event generator, since once the events are
generated, one can  add an additional
step and change  the energy sharing between the two nucleons.

In summary, we  extract the TE contribution by taking the difference between electron scattering data
and the predictions of the $\psi'$ formalism for  QE scattering. Therefore, predictions using ESF
for QE with the inclusion of the TE contribution fully describe electron scattering data by construction.

Including the TE model in neutrino Monte Carlo generators  is relatively simple.  The first step is to modify
the magnetic form factors for the proton and neutron as given in equation \ref{magnetic}. This accounts
for the  increase in the integrated  QE cross section. The second step is to change the relative faction
of the 1p1h and 2p2h process as given in equation \ref{eq-frac}, which changes shape of the
QE distribution in $\nu$.

  The  effective spectral function model and the TE model are not coupled. One
can use the effective spectral function to describe the scattering from independent
nucleons, and use another theoretical model  to account for the additional  
contribution from multi nucleon process.  Alternatively, one can use an alternative
model for the scattering from independent nucleons and use the TE model to account
for the additional contribution from  multi nucleon processes.

\section{Effective spectral functions for deuterium}
\label{deuteron-sub}
Neutrino charged current QE cross sections for deuterium are not modeled in current neutrino  Monte Carlo generators.  We find that 
neutrino interactions on deuterium can also be modeled with an    {\it{effective spectral function}}.

We use the theoretical calculations of reference \cite{D2} to predict the shape of the  transverse differential cross section
($\frac{1}{\sigma} \frac{d\sigma}{d\nu}(Q^2,\nu)$) for deuterium 
at several values of $Q^2$ as a function of $\Delta \nu$ = $\nu -Q^2/2M$.
 These theoretical  calculations are in agreement with electron scattering data.
We tune the parameters of the  {\it{effective spectral function}} to reproduce the spectra
predicted by the  theoretical calculations of reference \cite{D2}.

Fig. \ref{D-0.5} shows comparisons of model predictions for   $\frac{1}{\sigma} \frac{d\sigma}{d\nu}(Q^2,\nu)$  for QE 
electron scattering on deuterium  as a function
of $\Delta \nu$ = $\nu -Q^2/2M$ at $Q^2$=0.5 GeV$^2$.  The solid black line is the prediction from reference\cite{D2}.
 The red line is the prediction of the best fit parameters for the  {\it{effective spectral function}}  for deuterium.  For comparison we also show (in blue)  the prediction
for a Fermi gas  model with Fermi momentum $K_F$ = 0.100 GeV. 

For reference, we note that  the current default version of GENIE  cannot be used nuclei with atomic weight $A<7$.  This is because GENIE $K_F$ = 0.169 GeV (which has been extracted by Moniz\cite{moniz} for $\rm ^{3}Li_6$) for  all nuclei which have atomic weight $A<7$.  GENIE with the implementation of the  {\it{effective spectral function}} can be used for all nuclei.  Recently,  the {\it{effective spectral function}}  has been
implemented as an option in private  versions of NEUT\cite{callum} and GENIE\cite{Brian}.

  The best fit  parameters for the 
 {\it{effective spectral function}} for deuterium are given in Table \ref{fitsC}.  For deuterium, the 2p2h process 
is the only process that can happen.

\section{Pauli suppression}

The Pauli suppression in deuterium  is smaller than the  Pauli suppression in heavier nuclei.  The
multiplicative  Pauli suppression
factor  for  deuterium $K_{Pauli}^{deuteron}$ has been calculated by  S. K. Singh  \cite{dpauli}.  $ K_{Pauli}^{deuteron}$ 
can be parametrized\cite{dpfit}  by the following function.
\begin{eqnarray}
K_{Pauli}^{deuterium} &= & 1 - A e^{B(Q^2)^C}
\label{dpauli}
\end{eqnarray}
where, A=0.588918, B=-17.2306, C= 0.749157.   A comparison of
the Pauli suppression factors for $\carbon$ (eq.  \ref{nucpauli}) and deuterium (eq.  \ref{dpauli}) is shown
in Fig. \ref{DCpauli}.
\section{Conclusion}
We present parameters for an  {\it{effective spectral function}} that reproduce
the prediction for  
 $\frac{1}{\sigma} \frac{d\sigma}{d\nu}(Q^2,\nu)$ from the best
currently available models for  charged current QE
 scattering on nuclear targets.
We present parameters for a large number of nuclear targets
from  deuterium to lead.

  Since most of the currently 
available neutrino MC event generators model neutrino scattering in terms
of spectral functions,  the {\it{effective spectral function}} can easily be implemented.
For example, it has taken only a few days to implement
the    {\it{effective spectral function}} as an 
option in recent private  versions of NEUT\cite{callum} and GENIE\cite{Brian}.

The predictions using ESF
for QE with the inclusion of the TE contribution fully describe electron scattering data by construction.

\section{Acknowledgments}
We thank  Tomasz Golan for providing  us with the predictions of NuWro, and
Callum D. Wilkinson for implementation of the  {\it{effective spectral function}} in NEUT.

\section{Appendix A: Fermi smearing in the resonance region}
\subsection{ The method of Bosted and Mamyan}
Bosted and Mamyan\cite{super2} model Fermi motion effects for  electron scattering data in the resonance
and deep inelastic  region by smearing
the  structure function $W_1(W^\prime,Q^2)$  on free nucleons
to obtain the Fermi smeared structure function $W_1^F(W,Q^2)$. The smearing
is done over   $(W^\prime)^2$ (which is the square  of the mass of the hadronic final state)
at fixed values of $Q^2$.   Bosted and Mamyan\cite{super2} use the following
prescription.
\begin{eqnarray}
W_1^F(W^2,Q^2) &= &  \sum_{i=1~to~99} [
Z W_1^p((W_i^\prime)^2,Q^2) \nonumber \\
& +& (A-Z) W_1^n((W_i^\prime)^2,Q^2)] f_i(\xi_i)
\end{eqnarray}
where the sum approximates an integral. 
Here,  $W_1^p$ and $W_1^n$ are the free proton~\cite{Christy}
and neutron~\cite{Bosted} structure functions. 

 The shifted values $(W_i^\prime)^2$  are defined as
\begin{equation}
\label{shift}
(W_i^\prime)^2 = W^2 + \xi_i K_F |\vec q|
- 2 E_{\mathrm{shift}} (\nu + M)
\end{equation}
\noindent  
where $\xi=2k_z/K_F$, and $E_{\mathrm{shift}}$  is the 
energy shift parameter.  
 In the sum they use
99 values of $\xi_i$
\begin{equation}
\xi_i = -3+ 6(i-1)/98
\end{equation}
In the above equation  $f_i(\xi)$ is the normalized
probability for a nucleon to have 
a  fractional longitudinal momentum $\xi=2k_z/K_F$.
Bosted and Mamyan use the following  normalized
Gaussian for the  probability.
\begin{eqnarray}
P^{BM}(k_z) &=& P ^{BM}(\xi)= {\cal N}  (\xi=0, \sigma_{\xi}=1)
\end{eqnarray}
which is equivalent to
\begin{eqnarray}
f_i &=& 0.0245  e^{(-\xi_i^2/2)}
\end{eqnarray}
\noindent The  sum is a step-wise integration
over a Gaussian whose width is controlled by a Fermi
momentum $K_F$, truncated at $\pm 3\sigma$
($\xi_i$ ranges from -3 to +3), with
a shift in central $W$ related to the energy shift parameter
$E_{\mathrm{shift}}$. The values of $K_F$ and $E_{\mathrm{shift}}$ used for
the different nuclei are given in Table~\ref{tab:kfes}.
\begin{table*}
\begin{center}
\begin{tabular}{|c|c|c|c|c|c|c|c|c|c|} \hline 
Parameter  & $\rm ^{2}H$  & $\rm ^{3}He$    & $\rm ^{4}He$    &  $\rm ^{12}C$   & $\rm ^{20}Ne$   & $\rm ^{27}Al$  &  $\rm ^{40}Ar$ &  $\rm ^{56}Fe$
 &  $\rm ^{208}Pb$ \\ \hline\hline
 $\Delta$(MeV) &  0.13 &   5.3      &   14.0    &  12.5      & 16.6      & 12.5    & 20.6     & 15.1        & 18.8   \\ \hline
$f_{1p1h}$& 0 & 0.312    &  0.791   &    0.808  & 0.765    & 0.774  & 0.809    & 0.822     & 0.896  \\ \hline
$g_1$         & 0.2181 &   0.2560     &   0.1325      &  0.03819     & 0.1063    &0.1063  & 0.1019    & 0.07645     &  0.1474\\ \hline
$g_2$         &  0.6402 &  0.4343    &  0.3818       &   0.4168       & 0.3443 &  0.3443    & 0.3529   & 0.3892  & 0.3366  \\ \hline
$\sigma_1$ &  0.4465   &  0.7321     &   0.5724       &  0.3688    & 0.5538    &  0.5538   &  0.5401     & 0.5049     &  0.5964\\ \hline
$\sigma2$    & 0.9920 & 1.459       & 1.054 &         0.910           & 0.9555    &  1.008     & 0.9416      & 0.9356   &  0.9792\\ \hline
$\sigma_3$  &  3.233 &  3.126     & 2.122  &            1.928         & 1.870    &  1.935         &1.873     & 1.889 & 1.900 \\ \hline\hline
$K_F$(GeV)   &   0.100  &  0.115    &   0.190  &  0.228   &0.230 &  0.236  & 0.241   & 0.241     &  0.245\\ \hline
$a$             &  97.79 &  27.87  &   0 15.07   & 11.02      &  15.18   & 14.18  &  13.50  & 14.14&  13.96  \\ \hline
$b$            &  -0.2351 &  0.8118   & 0.6972  & 0.6772  & 0.5387   &  0.5466   &0.5471   & 0.5252   &  0.5131  \\ \hline
$c$             &  96.24  &  22.53 &  11.18        &  7.770  & 11.66 &     10.74 &   10.15  &      10.78 &  10.72 \\\hline
$d$             &  43.34 &   15.54  & 8.692         &  5.441    & 8.482  & 8.001  & 7.169 &       7.536 &  7.086     \\ \hline
\end{tabular}
\caption{ Top half: Parameters for the parameterization of the  one dimensional  projection along $k_z$ of  the  {\it{effective spectral function}}  ($P^{ESF}(\xi)$) for various nuclei.   The parameters for  deuterium  are shown in the column labeled  $\rm ^{2}H$ .  Bottom half:  Parameters for the parameterization of the the mean $<k^2>$ as a function of $k_z$. }
\label{fitsK}
\end{center}
\end{table*}
\begin{figure}
\begin{center}
\includegraphics[width=3.5in,height=2.3in]{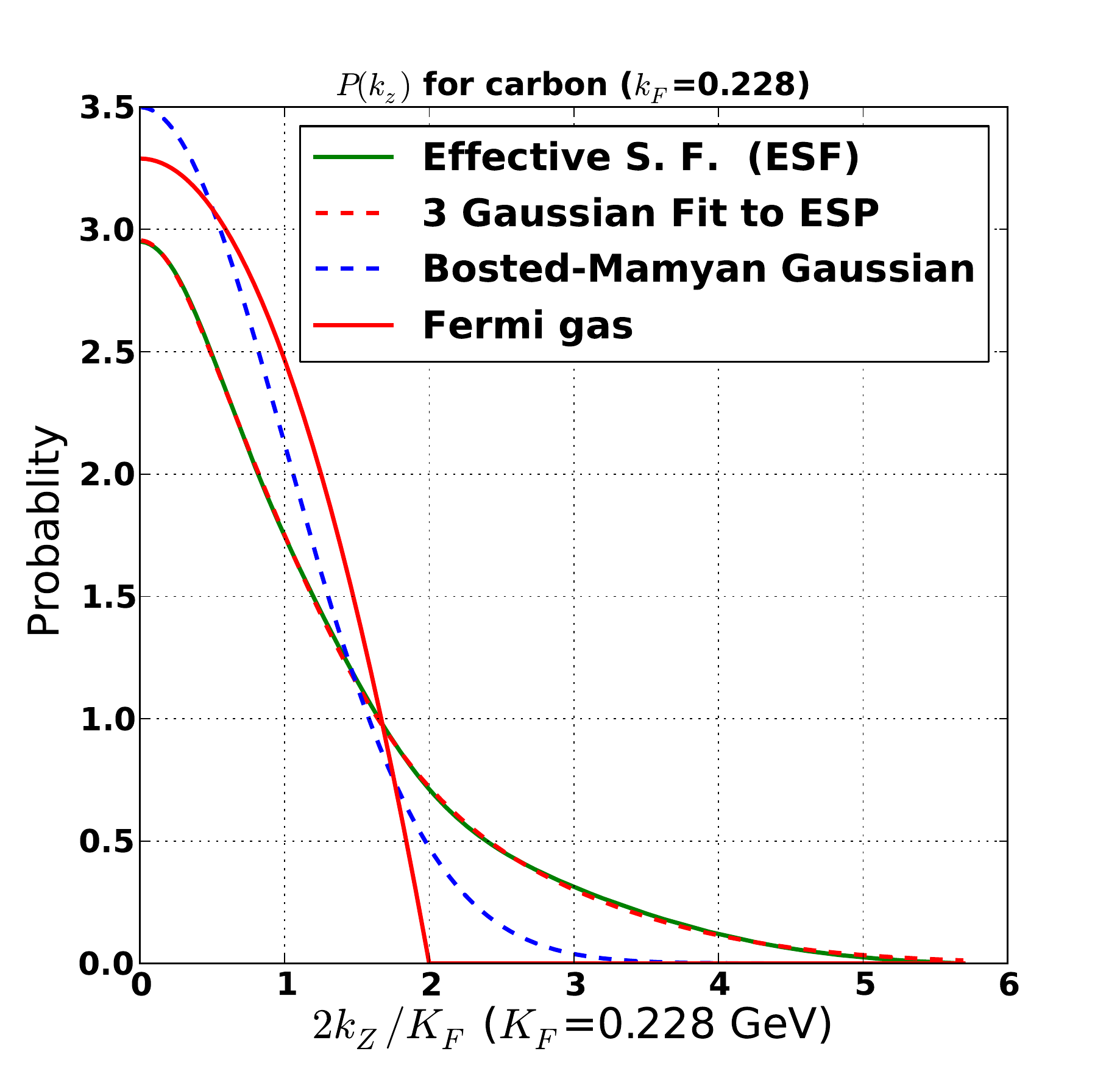}
\includegraphics[width=3.5in,height=2.3in]{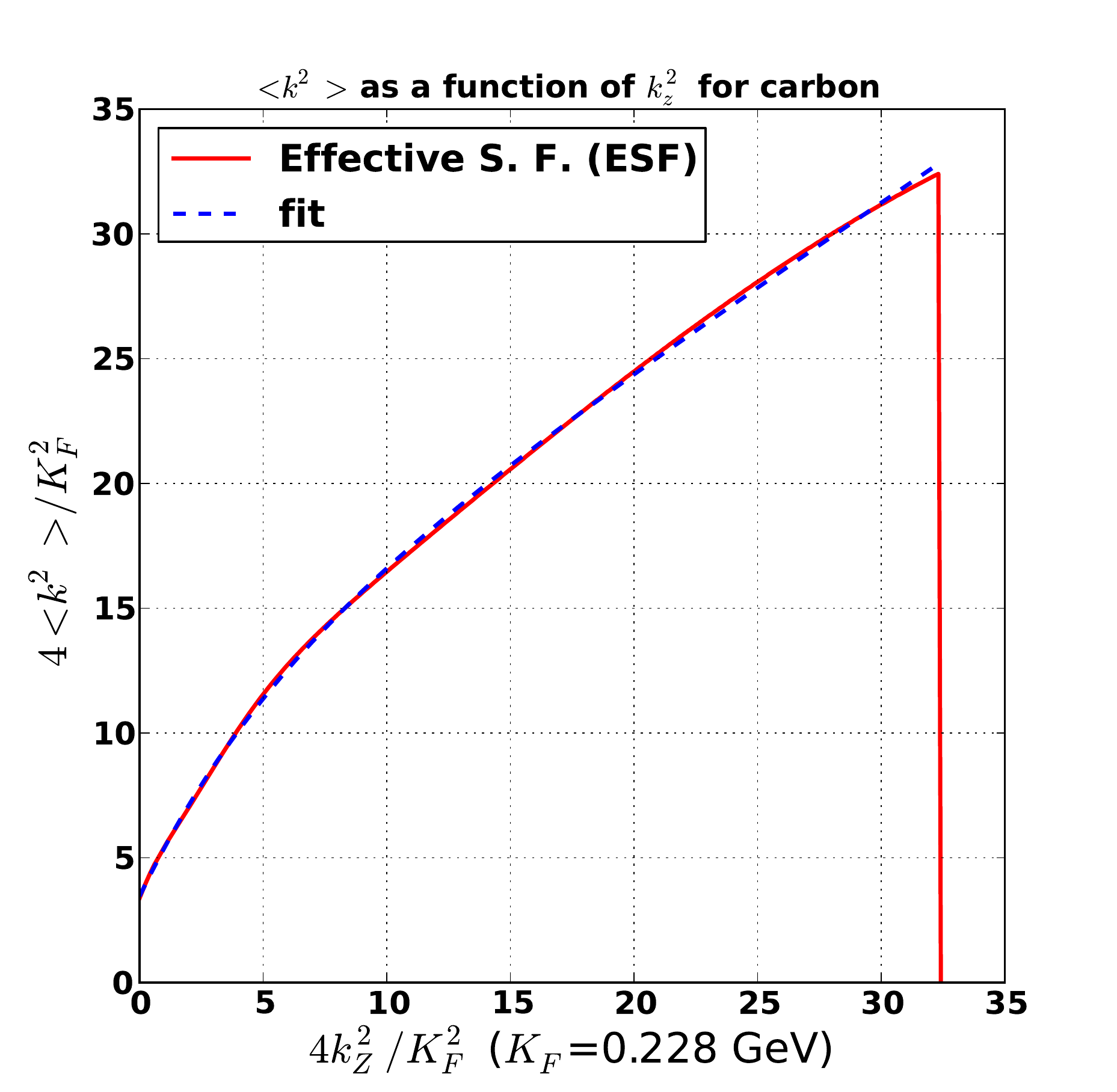}
\caption{ (a) A comparison of the probability distribution $P(k_z)$ for $\carbon$ plotted versus the
variable $\xi=2k_z/K_F$ for several spectral functions. (b) 
The average value of the square of  nucleon moment $k^2$ for the ESF for $\carbon$
versus the square of its $z$ component, $k_z^2$,  shown
in the form of $4 <k^2>/K_F^2$ versus $\xi^2=4k_z^2/K_F^2$. (Color online)      }
\label{kzfig}
\end{center}
\end{figure} 
%
%
%
\begin{figure}
\begin{center}
\includegraphics[width=3.5in,height=2.6in]{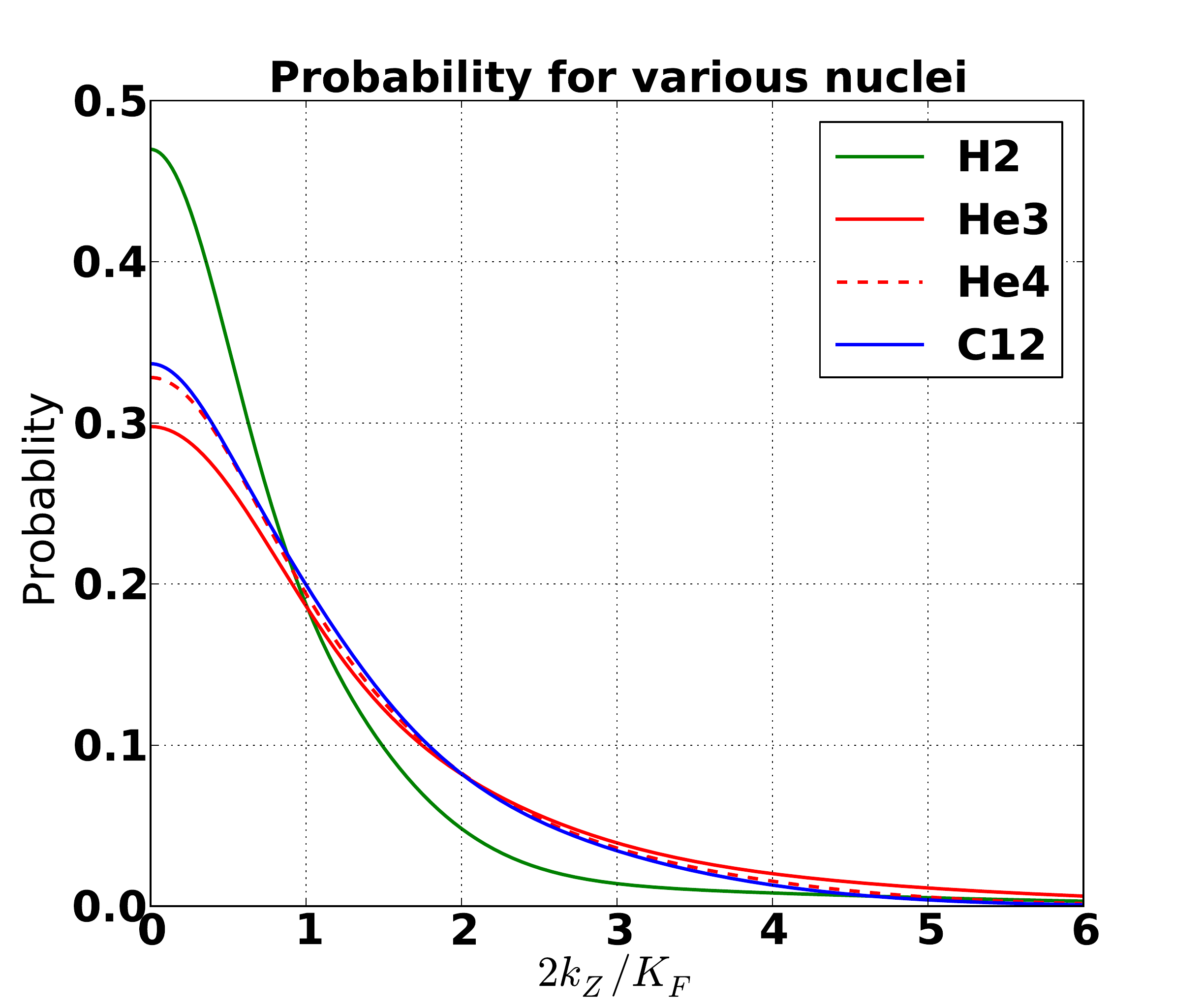}
\includegraphics[width=3.5in,height=2.6in]{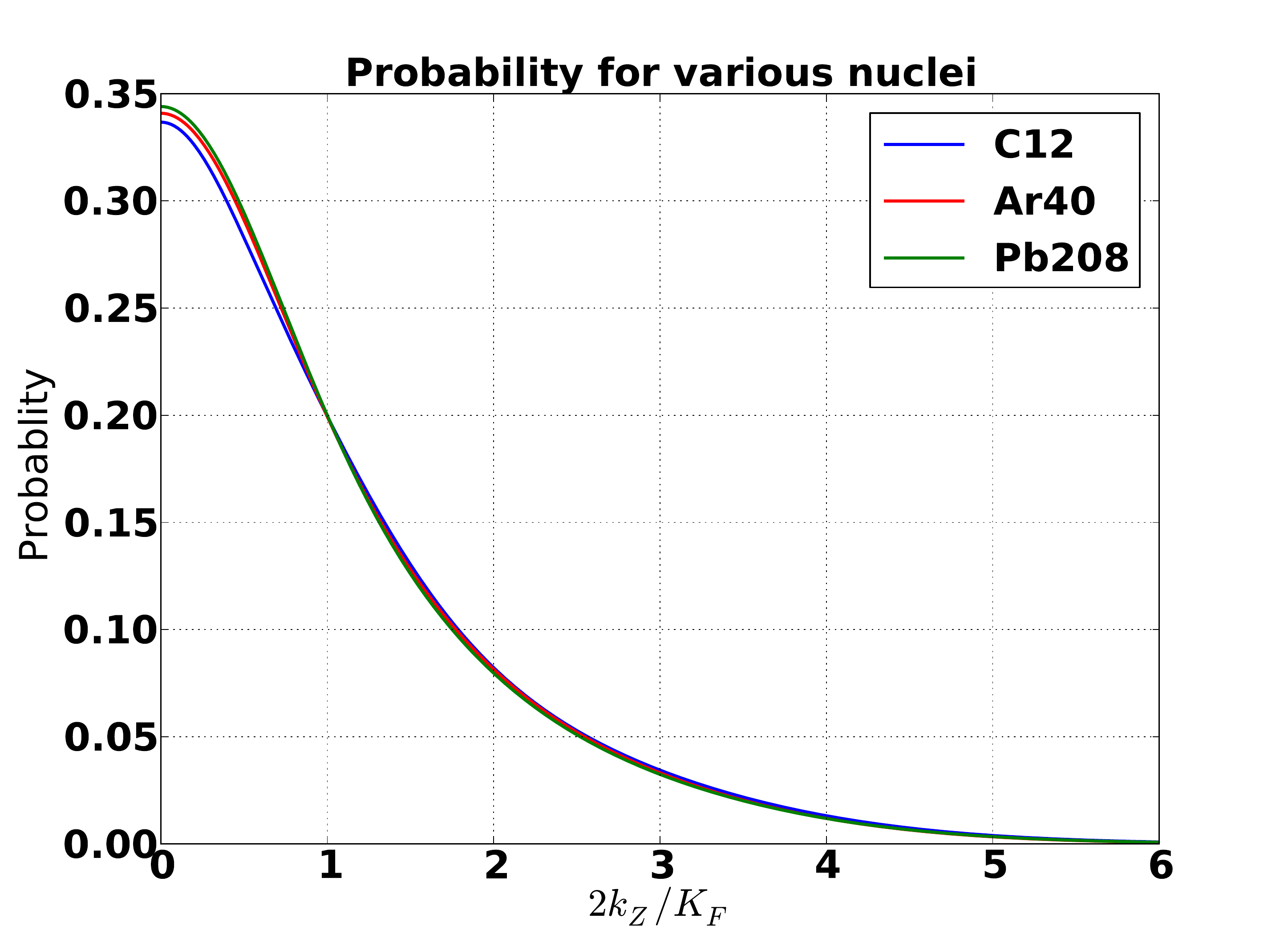}
\includegraphics[width=3.5in,height=2.6in]{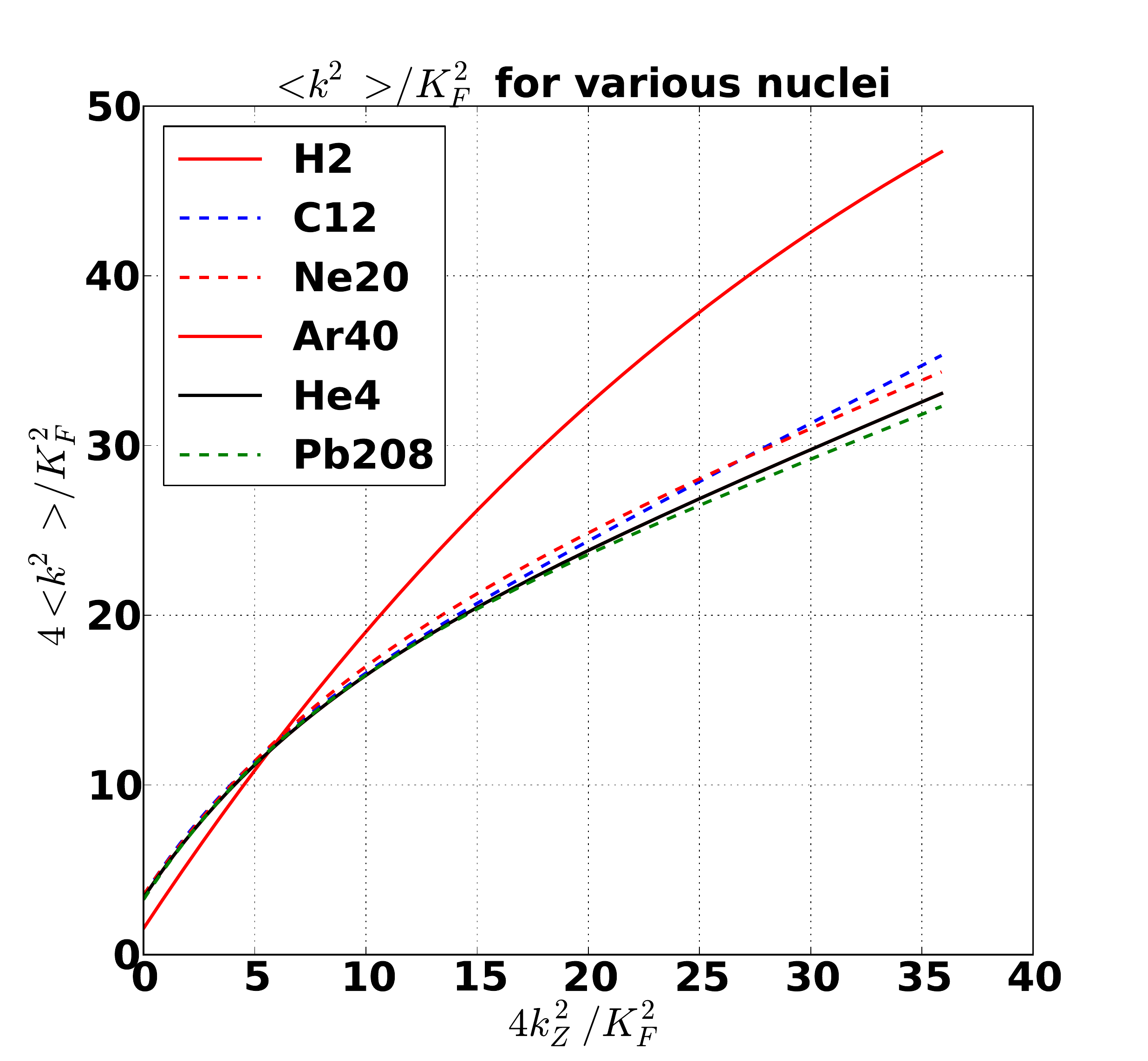}
\caption{ Top panel:  $P^{ESF}(\xi=2K_z/K_F)$ for light nuclei.  Middle panel: $P^{ESF}(\xi)$ for heavy nuclei.
Bottom panel: The mean $<k^2>$ as a function of $k_Z^2$ shown as $4<k^2>/K_F^2$ versus  ($\xi^2=4K_z^2/K_F^2$). 
Aside from deuterium (labeled H2) and Helium 3 (labeled He3) the functions for all nuclei
are similar.  Therefore, to a good approximation, the functions for $\carbon$ can be used all nuclei above A=12, and
also for for Helium 4  (provided that the appropriate Fermi momentum for each nucleus as  given in Table  \ref{fitsK} is used
for each nucleus)(color online).}
\label{multiNk2}
\end{center}
\end{figure} 
%

%
\subsection{Fermi smearing in the resonance region using the effective spectral function}
We have calculated the probability $P(k_z)$ for
the  {\it{effective spectral function}} ($P^{ESF}(k_z)$) for $\carbon$.
The top panel of Fig. \ref{kzfig} shows the probability
distributions  $P(k_z)$ plotted versus the
variable $\xi=2k_z/K_F$ for the  {\it{effective spectral function}}  as
compared to the distribution used by Bosted and Mamyan.
Also shown is the probability distribution for the Fermi Gas
model with $K_F$=0.228.  The  {\it{effective spectral function}} extends to higher momentum.  In order
to implement the  {\it{effective spectral function}}  we have
fit $P^{ESF}(k_z)$ to a sum of three normal Gaussians
with zero mean and different standard deviations
 $\sigma_{\xi}=\sigma_i$, and fractions  $g_1, g_2$, and
 $g_3=1-g_1-g_2$. 
\begin{eqnarray}
 P^{ESF}(\xi)&=& g_1 {\cal N} ( \sigma_1)+
g_2 {\cal N} ( \sigma_2) \nonumber\\
&+ &g_3 {\cal N} ( \sigma_3)\\
{\cal N} (\sigma_j)&=&  \frac {1}{\sigma_j \sqrt{2\pi}} e^{- 0.5\xi^2/\sigma_j^2}
\nonumber
\end{eqnarray}
where    $g_1=0.0382$, $g_2=0.417$, $\sigma_1 = 0.369$,  $\sigma_2 =0.910 $,
and  $\sigma_3 =1.928$.
For smearing with the $\xi$ distribution of the  {\it{effective spectral function}}  we
also use a 99 step integration in $\xi_i$ where,

\begin{eqnarray}
\xi_i &=& -6+ 12(i-1)/98 \nonumber \\
f_i (\xi)& =&\frac{g_1}{8.1585}{\cal N}  ( \sigma_1)+\frac{g_2}{8.1585} {\cal N} ( \sigma_2) \nonumber \\
&+ &\frac{g_3}{8.1585} {\cal N} ( \sigma_3)
\end{eqnarray}

Here, the  sum is a step-wise integration
over a Gaussian whose width is controlled by a Fermi
momentum $K_F$, truncated at $\pm 6\sigma$
($\xi_i$ ranges from -6 to +6), with
a shift in central $W$ related to removal energy. The values of $K_F$ used for
the different nuclei are given in Table~\ref{tab:kfes}.

Bosted and Mamyan calculate the shifted
values of $W'$ (equation \ref{shift}) using a  fixed  value for the energy shift $E_{\mathrm{shift}}$.
Instead, we calculate the shifted
values of $W'$ (equation \ref{WWW}) using the off-shell neutron energies ($E_n$) for  the
1p1h (eq. \ref{eq-W1p1h}) and 2p2h (eq.  \ref{eq-W1p1h}) processes, respectively.
In order calculate the  $k$-dependent off-shell neutrino energies  we need
to find the average $k^2$ as a function of $\xi$.

The bottom panel of Fig. \ref{kzfig} shows
the average value of the square of  nucleon moment $<k^2>$ 
versus the square of its $z$ component ($k_z^2$) calculated
for the  {\it{effective spectral function}} 
 for $\carbon$. What is shown specifically
is  $4 <k^2>/K_F^2$  as a function of $\xi^2=4k_z^2/K_F^2$.
We  parameterize ${<k^2>}$ by the following function:
\begin{eqnarray}
\label{k2-eq}
<k^2(\xi^2)> = \frac{K_F^2}{4} (a + b\xi^2-c e^{-\xi^2/d} )
\end{eqnarray}
%

We repeat the analysis for all other nuclei.  The top panel of Fig. \ref{multiNk2} shows
 $P^{ESF}(\xi=2K_z/K_F$) for light nuclei.  The  middle panel shows  $P^{ESF}(\xi)$) for heavy nuclei, and the
bottom panel shows the mean $<k^2>$ as a function of $k_Z^2$ presented as  as $4<k^2>/K_F^2$ versus  ($\xi^2=4K_z^2/K_F^2$). 

The parameters for  nuclei from $^{2}\rm H$  to $^{208}\rm Pb$ are  given in Table  \ref{fitsK}.
  However, we note that 
aside from $\rm ^{2}H$ (deuterium)  and $^3\rm He$,  the functions for all nuclei
are very similar.  Therefore,  as a good approximation, the parameters   for $ \carbon$ can be used  for all nuclei above A=12,
and also for $^4\rm He$ (provided that the appropriate Fermi momentum for each nucleus as  given in Table  \ref{fitsK} is used
for each nucleus).

%
\begin{figure}
\begin{center}
\includegraphics[width=3.5in,height=6.05in]{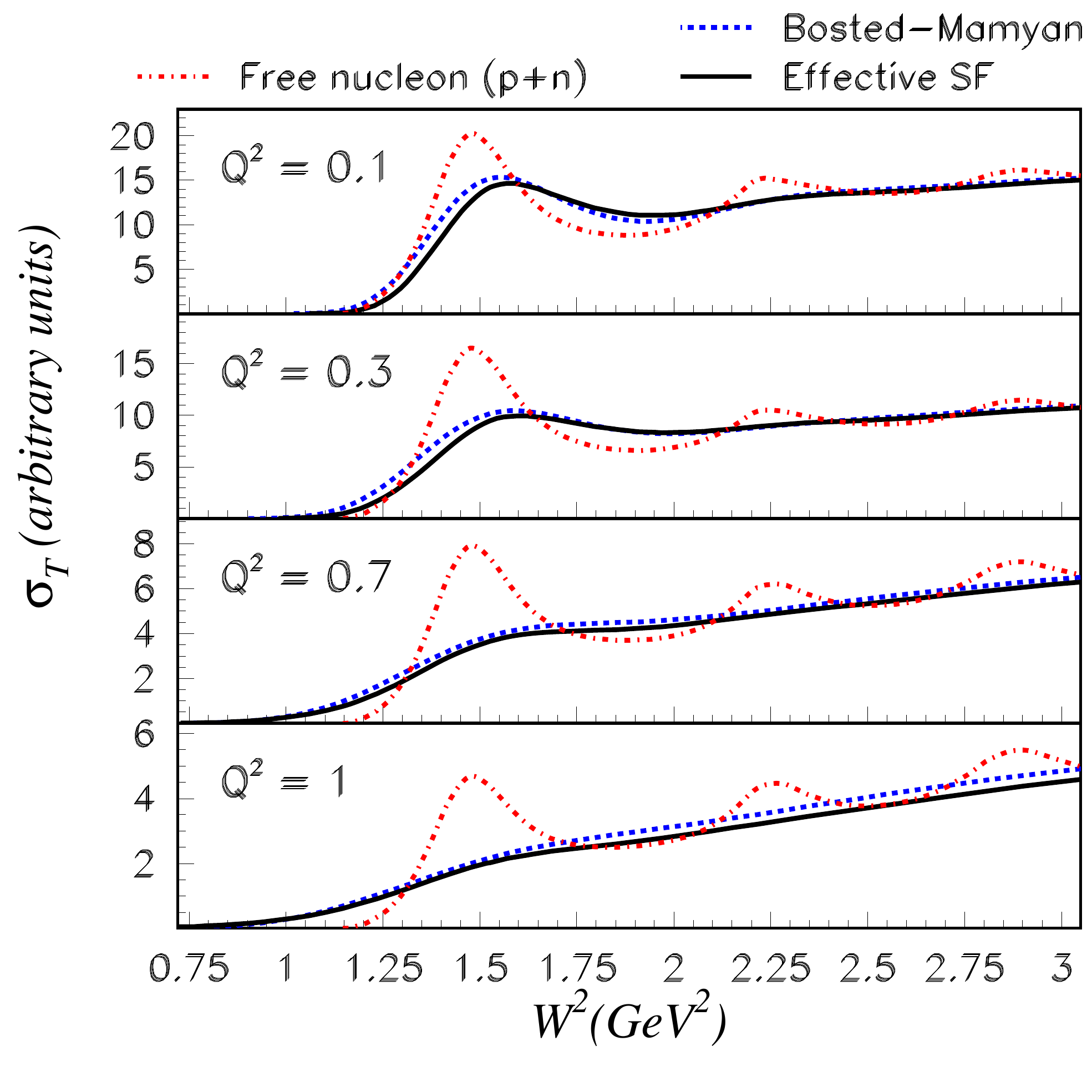}
\caption{  The results of Fermi motion smearing in $\carbon$ ($K_F=0.2280$ GeV)  
of the free nucleon cross sections in the resonance region using the   {\it{effective spectral function}}. 
 The results are shown for several values of $Q^2$ and
compared to the Fermi motion smearing used by Bosted and Mamyan 
 (color online).     }
\label{last}
\end{center}
\end{figure} 
%
%
%

Smearing with the  {\it{effective spectral function}}  requires that 
the  shifted value of  $W_i^\prime$ are different for the
the 1p1h and 2p2h contributions.
\begin{eqnarray}
W^2&=&M_n^2+2M_n\nu-Q^2\nonumber\\
(M_{n}')^2 &= & (E_n)^2 - {Vk^2} \nonumber\\
(W_i^\prime)^2&= &(M_{n}')^2 +2E_n\nu-\xi_i K_F |\vec q|- Q^2  \nonumber
\end{eqnarray}
Therefore, 
\begin{eqnarray}
\label{WWW} 
(W_i^\prime)^2 &=& W^2 -\xi_i K_F |\vec q| -V<k^2(\xi^2)> \nonumber\\
&+& [(E_n)^2-M_n^2] + 2\nu[E_n-M_n]   
\end{eqnarray}
Here  $<k^2(\xi^2)>$ is given by equation \ref{k2-eq}.

As mentioned earlier, the term $V(Q^2)$ multiplying $k^2$  in Equations \ref{eq-nu},  \ref{En1p1h},  \ref{En2p2h},  \ref{WWW}, \ref{eq-W1p1h},  and \ref{eq-W2p2h}  should be 1.0.  However, we find that in order to make the spectral function agree with $\psi'$ superscaling at very low $Q^2$ (e.g. $Q^2<0.3$~GeV$^2$) we need to apply a $Q^2$-dependent correction of the form $V=1-e^{-xQ^2}$ where $ x$=12.04 GeV$^{-2}$. This term, shown in Fig. \ref{vfactor}, accounts for the final state interaction (of the first kind) at low $Q^2$.  

For the 1p1h process  $E_n$ is given by equation \ref{En1p1h},
which when plugged into equation \ref{WWW} for
 $(W_i^\prime)^2$  yields the following expression: 
 \begin{eqnarray}
 \label{eq-W1p1h}
(W_i^\prime)^2_{1p1h}  &=& W^2 -\xi_i K_F |\vec q| -V<k^2(\xi^2)>    \\
&-&  2[\nu+M][\Delta+\frac{V<k^2(\xi^2)>}{2M_{A-1}^*} ]   \nonumber
\end{eqnarray}
 For the 2p2h process  $E_n$ is given by equation \ref{En2p2h} which
when plugged into equation  \ref{WWW}  for
 $(W_i^\prime)^2$  yields the following expression: 
 \begin{eqnarray}
 \label{eq-W2p2h} 
&(&W_i^\prime)^2_{2p2h}  = W^2 -\xi_i K_F |\vec q| -V<k^2(\xi^2)>    \\
&+&  [((M_p+M_n)-2\Delta -\sqrt{V<k^2(\xi^2)>+M_p^2} )^2-M_n^2] \nonumber \\ 
&+&  2 \nu [(M_p+M_n)-2\Delta -\sqrt{V<k^2(\xi^2)>+M_p^2} -M_n]\nonumber \
\end{eqnarray}
When smearing the proton and neutron structure functions,
the 1p1h and 2p2h processes are weighted by
the relative  fractions given in Table  \ref{fitsK}.

Fig. \ref{last}  shows the results of Fermi motion smearing in $\carbon$ ($K_F=0.2280$ GeV)  
of the free nucleon cross sections in the resonance region using the {\it{effective spectral function}}. 
 The results are shown for  several values of $Q^2$  and
compared to the Fermi motion smearing used by Bosted and Mamyan.  The spectra smeared
with the {\it{effective spectral function}} are
are shifter to higher values of $W^2$.

%

%
\section{Appendix B:  Calculation of the shape of the quasielastic  peak} 
We calculate the shape of the quasielastic  peak  $\frac{1}{\sigma} \frac{d\sigma}{d\nu}(Q^2,\nu)$ at fixed $Q^2$ using the
expressions below.
The on-shell  elastic $W_2$ structure function for
the scattering of neutrinos on free neutrons\cite{TE} ~is given by
\begin{eqnarray}
W_2^{on-shell}&=&  G (Q^2) \delta (\nu - Q^2/2M) 
\end{eqnarray}
where $G(Q^2)$  is given in terms of  vector and axial form factors.

\begin{eqnarray}
 G (Q^2) = |{ \cal F}_V (Q^2)|^2+|{\cal F}_A (Q^2)|^2\\ 
 |  {\cal F}_V(Q^2)|^{2}= \frac{[{\cal G}_E^V(Q^2)]^2+ \tau [{\cal G}_M^V(Q^2)]^2}{1+\tau}.
\end{eqnarray}
From conserved vector current (CVC)  $ {\cal G}_E^V(Q^2)$ and $ {\cal G}_M^V(Q^2)$ 
are related to the electron scattering electromagnetic form factors\cite{BBBA}
$G_E^p(Q^2)$, $G_E^n(Q^2)$, $G_M^p(Q^2)$, and $G_M^n(Q^2)$:
$$ 
{\cal G}_E^V(Q^2)=G_E^p(Q^2)-G_E^n(Q^2), 
$$
$$
{\cal G}_M^V(Q^2)=G_M^p(Q^2)-G_M^n(Q^2). 
$$
The axial form
factor ${\cal F}_A$  can be approximated by the dipole form
$$
{\cal F}_A(Q^2)=\frac{g_A}{\left(1+\frac{\D Q^2}{{\cal M}_A^2}\right)^2 },
$$
Where $g_A=-1.267$, and  ${\cal M}_A =1.014$~GeV\cite{BBBA}.
Fits  that include modifications to dipole form for both vector and  axial form
factors can be found in Ref.\cite{BBBA}.

We note that when we calculate the shape of the
QE peak $\frac{1}{\sigma} \frac{d\sigma}{d\nu}(Q^2,\nu)$ at fixed $Q^2$
  the function $G (Q^2)$  cancels out.

For the scattering from an off-shell nucleon, the 
energy conservation $\delta$ function
takes the following form:
\begin{eqnarray}
M_p^2&= &(M_{n}')^2 + 2E_n\nu-2 |q|  k_z- Q^2 \nonumber \\
\nu&=&  \frac {Q^2+2 |q| k_z + M_p^2 - (M_{n}')^2} {2E_n } \nonumber \\
W_2^{off}&=&  G (Q^2) \delta (\nu -  \frac {Q^2+2 |q| k_z + M_p^2 - (M_{n}')^2}{2E_n })
\label{delta}
\end{eqnarray}
where $k_z=k\cos\theta$.  Here $\theta$ is the angle between the direction
of the momentum transfer $\vec {q}$ and $\vec{k}$.
The shape of the quasielastic  peak $\frac{1}{\sigma} \frac{d\sigma}{d\nu}(Q^2,\nu)$ at fixed $Q^2$ is then given by
\begin{eqnarray}
\frac{d\sigma}{d\nu}  (Q^2,\nu)&\propto & \int_{0}^{K_M} 2\pi  \int_{-1}^{1}  W_2^{off} |\phi(k)|^2 k^2~d\cos\theta ~dk\nonumber\\
\label{integral}
\end {eqnarray}
Where    $P(k) = |\phi(k)|^2  4\pi k^2 $ is the probability
distribution for a nucleon to have
a momentum $k=|\vec{k}|$ in the nucleus. 
For all of the momentum distributions  that we investigate we set the probability to
zero for $k>K_M$ where  $K_M$=0.65 GeV.
%
%
%
%
%
%
\subsection{The shape of the quasielastic peak for the Fermi gas model}
For the Fermi gas  model we can get an approximate distribution for  $\frac{d\sigma}{d\nu} (Q^2)$ in closed form.  We
use this calculation as a check on our results which are obtained using 
equation \ref{integral}. 

For the Fermi gas model we do the calculation in  cylindrical coordinate 
$$(2\pi k^2~d\cos\theta ~dk =  \pi dk_r^2~ dk_z)$$
$$k=\sqrt{k_r^2+k_z^2}.$$
Equation \ref{integral} can then be written as
\begin{eqnarray}
  W_2^{off}&\propto& \frac{E_n}{|\vec q |}  \delta ( k_z - \frac{1}{2|\vec q|}  ( 2E_n \nu  - Q^2 - M_p^2 + (M_{n}')^2 ) \nonumber \\
\frac{d\sigma}{d\nu}  (Q^2)&\propto & \int_{-K_F}^{K_F}  \int_{0}^{{K_F-k_z^2}} W_2^{off} |\phi(k)|^2  \pi~ dk_r^2~ dk_z 
\end{eqnarray}
For the Fermi gas model the   momentum distribution is zero for   $k>K_F$, and  for $k<K_F$ it is given by
\begin{eqnarray}
|\phi(k)|^2&=& \frac {1}{N} 
~~~~~~~~N= \frac {4}{3} {\pi }  K_{F}^{3}  \nonumber \\
P(k)dk &=&  |\phi(k)|^2 4\pi k^2 dk   =  \frac {1}{N} {4\pi k^2 dk} 
\label{eqFermi}
\end{eqnarray}
For simplicity, we assume
that the energy of the off shell neutron is a constant which is independent
of k. Using  $<k^2> = \frac{3}{5} K_F^2$ we obtain
\begin{eqnarray}
<E_n>&= & M_n-\Delta-\frac{3K^2_F}{10M_{A-1}^*} \nonumber\\
<(M_{n}')^2>&=&<E_n>^2-3K_F^2/5
\label{approx}
\end{eqnarray}
\begin{eqnarray}
\frac{d\sigma}{d\nu}  (Q^2)&\propto & \int_{-K_F}^{K_F}  \int_{0}^{{K_F-k_z^2}} W_2^{off} |\phi(k)|^2  \pi~ dk_r^2~ dk_z   \nonumber \\ 
         &= & \int_{-K_F}^{K_F}  W_2^{off} \frac {3}{4K_F^3} (K_F^2-k_z^2)~ dk_z    
\end{eqnarray}
Integrating the $\delta $ function  in equation \ref{delta}  over $k_z$ we get
$$k_z = \frac{1}{2|\vec q|}  [ 2<E_n > \nu  - Q^2 - M_p^2 + <(M_{n}')^2> ]$$
\begin{eqnarray}
\frac{1}{\sigma} \frac{d\sigma}{d\nu}  (Q^2,\nu) &= &  \frac{<E_n>}{|\vec q|} \frac{3}{4} \frac {(K_F^2-k_z^2)}{K_F^3}.
\label{nuFermi}
\end{eqnarray}
The above equation satisfies the normalization condition
\begin{eqnarray}
\int_{}^{}  \frac{<E_n>}{|\vec q|} \frac{3}{4} \frac {(K_F^2-k_z^2)}{K_F^3} d\nu  &=& 1
\nonumber 
\end{eqnarray}


\begin{thebibliography}{9}
\bibitem{genie}  C. Andreopoulos [GENIE Collaboration], Acta Phys. Polon. B 40, 2461(2009), C.Andreopoulos (GENIE), Nucl. Instrum. Meth.A614, 87,2010.
\bibitem{neugen}  H. Gallagher, (NEUGEN) Nucl. Phys. Proc. Suppl. 112 (2002).
\bibitem{neut}  Y. Hayato (NEUT), Nucl Phys. Proc. Suppl.. 112, 171 (2002).
\bibitem{nuance}  D. Casper (NUANCE) , Nucl. Phys. Proc. Suppl. 112, 161 (2002) (http://nuint.ps.uci.edu/nuance/).
\bibitem{nuwro}  J. Sobczyk (NuWro), PoS NUFACT 08, 141 (2008),  C. Juszczak, Acta Phys. Polon. B40 (2009) 2507 (http://borg.ift.uni.wroc.pl/nuwro/).
\bibitem{gibuu}T. Leitner, O. Buss, L. Alvarez-Ruso, U. Mosel (GiBUU), Phys. Rev. C79, 034601 (2009) (arXiv:0812.0587).
\bibitem{moniz} E. J. Moniz et al. Phys. Rev. Lett. 26, 445(1971);  E. J. Moniz, Phys. Rev. 184, 1154 (1969);  R. A. Smith and E. J. Moniz, Nucl. Phys. B43, 605 (1972).
%
\bibitem{bf}  O. Benhar and S. Fantoni and G. Lykasov, Eur Phys. J.  A 7 (2000), 3, 415;  O, Benhar et al., Phys. Rev. C55. 244 (1997).
\bibitem{super1}  J.E. Amaro, M.B. Barbaro, J.A. Caballero, 
T.W. Donnelly, A. Molinari, and I. Sick, Phys. Rev. C 71, 015501 (2005). arXiv:nucl-th/0409078.
\bibitem{super2} P. E. Bosted, V. Mamyan, arXiv:1203.2262.
\bibitem{eric} M. E. Christy, private communication (2014).
\bibitem{Bodek-Ritchie} A. Bodek, and J. L. Ritchie, Phys. Rev. D23, 1070 (1980).  A. Bodek and J. L. Ritchie, Phys. Rev. D24, 1400 (1981).
%
\bibitem{bfit} S. Mishra, (NOMAD collaboration) private communication, fit parameters from  E. Iacopini  (1997). 
\bibitem{D2} M. E. Christy, N. Kalantarians, J. J. Ethier, and W. Melnitchouk. In preparation (2014).
\bibitem{dpauli} S.K.~Singh, Nucl. Phys. B36 419(1972);.K.~Singh and  H. Arenhove, 
\bibitem{dpfit}H. Budd, A. Bodek, J. Arrington, Nucl.Phys.Proc.Suppl. 139, 90(2005), Z. Phys. A 324, 3476 (1986).
\bibitem{Bosted} P.E. Bosted and M.E. Christy, 
Phys. Rev. C 77, 065206 (2008). (arXiv:0711.0159).
\bibitem{Christy}
M.E. Christy and P.E. Bosted,  Phys. Rev. C 81, 055213 (2010). (arXiv:0712.3731).
\bibitem{TE}  A. Bodek, H. S. Budd and M. E. Christy, Eur. Phys. J. C 71, 1726 (2011).
\bibitem{MEC4}
J. Carlson, J. Jourdan, R. Schiavilla, I. Sick, Phys.Rev. C65,  024002 (2002).
\bibitem{JUPITER}  A. Bodek, C.Keppel and M. E. Christy (spokespersons JUPITER Collaboration), Jefferson Lab 
Experiment E04-001.
\bibitem{callum} Callum D. Wilkinson, private communication (NEUT).
\bibitem{Brian} Brian Coopersmith, private communication (GENIE).
\bibitem{BBBA} 
A. Bodek, S. Avvakumov, R. Bradford, and  H. Budd, Eur. Phys. J. C{\bf
53}, 349 (2008). 
\end{thebibliography}
\end{document}